\documentclass{aa}  

\usepackage{graphicx}
\usepackage{txfonts}
\usepackage{tikz}

\usepackage{ulem}   
\usepackage{multirow}
\usepackage{float}

\usepackage{enumitem}

\usepackage[colorlinks=true,     linkcolor=blue, citecolor=blue, filecolor=blue, urlcolor=blue]{hyperref}

\definecolor{gr}{rgb}{0.2, 0.20, 0.80}

\begin{document}

\title{X-ray luminosity-star formation rate scaling relation: Constraints from the eROSITA Final Equatorial Depth Survey (eFEDS)}

\titlerunning{X-ray luminosity - star formation rate scaling relation: constraints from eFEDS}

\author{G. Riccio\inst{1}, G. Yang\inst{2,3} \and K. Ma\l ek\inst{1,4} \and M. Boquien\inst{5} \and Junais\inst{1} \and F. Pistis\inst{1} \and M. Hamed\inst{1} \and M. Grespan\inst{1}, M. Paolillo\inst{6,7,8} \and O. Torbaniuk\inst{9,10}  }

\institute{National Centre for Nuclear      Research, ul. Pasteura 7, 02-093        Warszawa, Poland
\and
Kapteyn Astronomical Institute, University of Groningen, P.O. Box 800, 9700 AV Groningen, The Netherlands
\and
SRON Netherlands Institute for Space Research, Postbus 800, 9700 AV Groningen, The Netherlands
\and
    Aix Marseille Univ. CNRS, CNES, LAM, Marseille, France
\and
    Instituto de Alta Investigación, Universidad de Tarapacá, Casilla 7D, Arica, Chile
\and
    Department of Physics, University of Naples Federico II, C.U. Monte Sant'Angelo, Via Cinthia, I-80126 Naples, Italy
\and
   INFN - Sezione di Napoli, 80126 Naples, Italy
\and
   INAF-Osservatorio astronomico di Capodimonte, Via Moiariello 16, I-30131 Naples, Italy
\and
    Department of Physics and Astronomy, University of Bologna, via Piero Gobetti 93/2, I-40129 Bologna, Italy
\and
    INAF - Osservatorio di Astrofisica e Scienza dello Spazio di Bologna, Via Gobetti 93/3, I-40129 Bologna, Italy}

  \abstract
   {}
   { We present measurements of the relation between X-ray luminosity and star formation activity for a sample of normal galaxies spanning the redshift range between 0 and 0.25. We use data acquired by SRG/eROSITA for the performance and verification phase program called eROSITA Final Equatorial Depth Survey (eFEDS). The eFEDS galaxies are observed in the 0.2-2.3 keV band.}
   {Making use of a wide range of  ancillary data, spanning from the  ultraviolet (UV) to mid-infrared wavelengths (MIR), we estimated the star formation rate (SFR) and stellar mass ($M_{star}$)  of 888  galaxies, using Code Investigating GALaxy Emission (CIGALE). In order to study sources whose X-ray emission is dominated by X-ray binaries (XRBs), we classified these galaxies into normal galaxies and active galactic nuclei (AGNs) by making use of the observed fluxes in the X-ray, optical, and MIR ranges, as well as the results from the SED fitting. To isolate the contribution of XRBs, which scale with the SFR and $M_{star}$, we subtracted the contribution of hot gas, coronally active binaries, and cataclysmic variables to the total X-ray emission. We divided our sample of normal galaxies in star-forming (SFGs) and quiescent galaxies according to their position on the main sequence.}
   {We confirm a linear correlation between the X-ray luminosity and the SFR for our sample of SFGs, shown previously in the literature. However, we find this relation to be strongly biased by the completeness limit of the eFEDS survey. Correcting for completeness, we find the fitted relation to be consistent with the literature. We also investigated the relation between X-ray emission from both LMXBs and HMXBs populations with $M_{star}$ and SFR, respectively. Correcting for completeness, we find our fitted relation to considerably scatter from the literature relation at high specific SFR ($SFR/M_{star}$). We conclude that without accounting for X-ray non-detections, it is not possible to employ eFEDS data to study the redshift evolution of the LMXBs and HMXBs contributions due to completeness issues. Furthermore, we find our sources to largely scatter from the expected Lx/SFR vs specific SFR relation at high redshift. We discuss the dependence of the scatter on the stellar mass, metallicity, or the globular cluster content of the galaxy.  }
    {}

\maketitle

\section{Introduction}
\label{sec:Introduction}

In recent decades, one of the main efforts of the high-energy astronomy community has been focused on calibrating the X-ray emission and source populations of galaxies against the star formation rate (SFR) and the stellar mass ($M_{star}$) (\citealt{Gilfanov2004, Mineo2014, Basu2013a, Lehmer2016}).
X-ray binaries (XRBs), the hot ionized interstellar medium (ISM), and active galactic nuclei (AGNs) are the main contributors to the total X-ray emission of galaxies. In partiicular, XRBs are stellar systems composed of an extremely dense object (a neutron star or black hole) that accretes mass from a secondary star. They can be divided into two main categories: high-mass X-ray binaries (HMXBs), when the donor star is an early-type star (OB star, or a supergiant), or low-mass X-ray binaries (LMXBs) when the secondary star is a later-type star (typically of M, K spectral types). It is well known that the X-ray emission from these objects traces the stellar population of the galaxy. Indeed, the number of HMXBs and their collective X-ray luminosity was found to scale with the star formation rate (SFR) of the host galaxy (\citealt{Grimm2003}; \citealt{Lehmer2010}; \citealt{Mineo2012}, \citealt{Mineo2014}).  This fact is well understood in terms of the short evolutionary time scales of HMXBs and by the fact that the secondary star is a young supergiant (e.g., \citealt{Verbunt1995}, \citealt{Gilfanov2007}). On the other hand, the number of LMXBs and their X-ray luminosity are correlated with the $M_{star}$ of the galaxy. Also, joint relations were found between the X-ray luminosity, SFR, and the  $M_{star}$  (\citealt{Lehmer2016}). In their study, \cite{Fragos2013} employed local scaling relations data to restrict the predictions of theoretical XRB population-synthesis models. They found that the spectral energy distribution (hereafter, SED) of the XRBs remains relatively unchanged with redshift, despite a substantial evolution of its normalization, which occurs primarily as a result of changes in the cosmic SFR. However, the particular X-ray output of XRBs is affected by metallicity and mean stellar age. In particular, the X-ray luminosity per unit of star-formation rate from HMXBs varies by  order of magnitude when moving from solar metallicity to metallicity below 10\%, while the X-ray luminosity per unit of stellar mass from LMXBs reaches a peak at the age of around 300 million years and then gradually decreases at later times (see Fig. 2 of \citealt{Fragos2013}). For mean stellar ages exceeding approximately 3 billion years, there is little variation in the X-ray luminosity from LMXBs. These relations provide analytical and tabulated guidelines for the energy output of XRBs, which can be directly integrated into cosmological simulations or models of the X-ray emission of galaxies.

The X-ray luminosity-SFR relation provides estimates of the SFR that are less affected by uncertainties due to dust and gas absorption, as the X-ray light is less affected by interstellar extinction than other traditional indicators. This characteristic makes it a valuable tool for cross-calibrating various SFR indicators and diagnosing star formation in galaxies.
However, when observing distant galaxies, distinguishing the emission from  HMXBs, LMXBs, or supermassive black holes is challenging. In fact, 
AGNs seem to dominate the total X-ray emission of bright galaxies \citep{Xue2011, Lehmer2012}, and the latest studies suggest the presence of co-evolution or relations between AGN and star-formation activity of the host galaxy \citep{Aird2019, Torbaniuk2021}.
 Also, it is known that galaxies contain a considerable amount of ionized gas at temperatures of around $10^7-10^8$K, which contributes to the total X-ray emission \citep{Grimes2005, Tullmann2006}. This hot gas can form in different ways, from mass loss of old stellar populations (e.g., stellar winds from
evolved stars, planetary nebulae, and Type Ia supernovae) and accretion of the
intergalactic medium, as well as mergers of small galaxies and can dominate the X-ray emission of the galaxy in the soft band (0.5-2 keV). As only the XRB emission correlates with stellar population properties, it is an extremely difficult challenge to disentangle the contribution of each of these components to probe the aforementioned scaling relations.

In this paper,  we make use of the eROSITA Final Equatorial Depth Survey (eFEDS, \citealt{Brunner2022}) to study the correlation between SFR and X-ray luminosity.
The extended ROentgen Survey with an Imaging Telescope Array(eROSITA; \citealt{Merloni2012}, \citealt{Merloni2020}, \citealt{Predehl2021}) as part of the Spectrum-Roentgen-Gamma (SRG, \citealt{Sunyaev2021}) mission, has become a crucial tool for investigating the X-ray characteristics of galaxies.
The X-ray observations are combined with UV (GALEX), optical and near-IR (KiDS, HSC, VISTA/VHS), and mid-infrared (WISE) data (\citealt{Salvato2022}). These data are used to fit the SED of the galaxies to estimate their physical properties. The paper is organized as follows. In Section~\ref{Sec:Data}, we describe the dataset and the sample selection based on the quality of the photometry. In Section~\ref{Sec:Methodology} we present the broadband SED fitting method and the first AGN selection based on photometry. In Section~\ref{sec:reliability_check}, we discuss the reliability of the estimated physical properties. 
We describe the subtraction of other X-ray components, such as hot gas, and a second AGN selection based on the SED fitting in Section~\ref{sec:substraction}.
 We discuss the results on the Lx-SFR and Lx-sSFR relations in Section~\ref{Sec:Results}. Our summary and conclusions are presented in Section~\ref{Sec:Conclusions}. 
Throughout this paper we use the WMAP7 cosmology \citep{Komatsu2011}: $\Omega_m = 0.272$,
$\Omega_{\Lambda}$ = 0.728, and $H_0$ = 70.4 km s$^{-1}$ Mpc$^{-1}$.

\section{Data and sample selection}
\label{Sec:Data}

This work is based on a combination of eROSITA and ancillary data spanning from the ultraviolet (UV) to the mid-infrared (MIR). Here, we briefly summarize the main properties of the catalog and refer the reader to the cited works for more details.

The X-ray sources presented in this work have been detected by eROSITA, the primary instrument aboard the SRG orbital observatory (\citealt{Sunyaev2021}). The main objective of the SRG mission is to perform a four-year survey of the full sky in continuous scanning mode. The sources taken into account in this work are part of the eFEDS, which scans $\sim$ 140 square degrees of the sky as a  verification phase ahead of the planned four years of all-sky scanning operations. With the exception of  the all-sky surveys, eFEDS represents the largest contiguous X-ray survey in the soft X-ray band. A detailed explanation of the data processing and properties of the catalog can be found in \cite{Brunner2022}. The catalog includes 27\,910 X-ray sources detected in the band 0.2-2.3 keV, with detection likelihoods of $\geq6$, corresponding to a (point source) flux limit of $6.5 \times 10^{-15}$ erg $s^{-1}cm^{-2}$ in the 0.5–2.0 keV energy band. To ensure the highest signal-to-noise ratio (S/N) of the X-ray sample used in this work, we excluded sources located at the border of the fields (\textsc{inArea90} flag, 3\% of the total).

In order to estimate galaxies' physical parameters, multi-wavelength observations are required. \cite{Salvato2022} provides a catalog of multi-wavelength counterparts and redshifts of the X-ray sources presented in \cite{Brunner2022}. Considering the large PSF ($\sim 16\arcsec$) and the small number of photons associated with a typical X-ray detection, the positional uncertainties of the sources can be in the order of the arcsecond, making the identification of the counterpart impossible to determine by the closest neighbor match alone. To overcome this problem \cite{Salvato2022} performed two different methods specifically developed to identify the correct counterparts to X-ray sources: 1) \textit{NWAY} (\citealt{Salvato2018}), based on Bayesian statistics; 2) \textit{ASTROMATCH} (\citealt{Ruiz2018}), based on maximum likelihood ratio (\citealt{Sutherland1992}). A detailed description of these methods can be found in \cite{Salvato2022}. The DESI Legacy Imaging Survey DR8 (LS8; \citealt{Dey2019}) was used for the optical counterpart identification. Together with the LS8 data, the UV, optical, and infrared photometry was also included. 

Each counterpart is assigned a quality flag, \textsc{CTP\_quality}, that characterizes the quality of the cross-match. To ensure reliable optical photometry for the SED fitting, we selected only those objects having \textsc{CTP\_quality}>2, which are objects for which both methods are in agreement with respect to the counterpart, but only one assigns to it a cross-match probability above the threshold (\textsc{CTP\_quality}=3) or both agree on the counterpart and have an assigned probability above the threshold (\textsc{CTP\_quality=4}). In this way, we selected  22\,256 objects (81\% of the total sample). 

Sources are classified as Galactic or extra-Galactic (see details in Section 5 of \citealt{Salvato2022}). To ensure the removal of foreground Galactic stars from the analysis, we selected only those objects flagged as \textsc{SECURE EXTRAGALACTIC} (5100, 19\% of the total sample). Furthermore, a redshift quality flag is given to the sources, \textsc{CTP\_REDSHIFT\_GRADE}, in a range from 5 (spectroscopy) to 0 (unreliable photo-z). Photometric redshifts are computed using LePHARE code (\citealt{Arnouts1999}. \citealt{Ilbert2006}) and the estimates were then compared with those obtained with DNNz (Nishizawa et al. in prep.), a machine learning method that uses HSC photometry. A detailed description of the method can be found in \cite{Salvato2022}. In this work, we adopt the selection criteria \textsc{CTP\_REDSHIFT\_GRADE} $\geq3$, which includes all previously selected sources (5100, 19\% of the total sample).

The main goal of this work is to analyze the properties of the eROSITA sample of normal galaxies (non-AGN systems) that are expected to have relatively low X-ray luminosities ($L_x\leq 3\times 10^{42}$ erg/s) compared to an AGN system (\citealt{Luo2017}, \citealt{Lehmer2016}). Such systems are hardly observed at high redshift, due to the limit on the sensitivity of the instrument. Taking into account the eFEDS sensitivity limit, we restrict our analysis to the sources having $z<0.35$ (888, 3.2\% of the total sample).

In our analysis, the galaxies' physical properties are estimated via an SED fitting, which requires high-quality multi-wavelength measurements to ensure reliable results. In particular, the SFR requires high-quality IR observations to account for the amount of UV light absorbed and re-emitted by the dust. Therefore, we require the X-ray sources to be observed in WISE1 and WISE2 bands with signal-to-noise $S/N\geq2$. The previously selected sample (888, 3.2\% of the total sample) fulfilled this criterion. We stress that all the sources selected with these requirements are also characterized with $S/N\geq2$ for all the other available photometry, with the exception of W3 and W4 bands.
At the end of the selection process, we restrict our sample to 3\% of the initial \cite{Salvato2022} catalog. The selection process ensures the quality and reliability of the SED fitting procedure, explained in the next section. Table~\ref{table:1} shows the ancillary data associated with the X-ray sources. We stress that all the selected sources have reliable spectroscopic redshift (\textsc{CTP\_REDSHIFT\_QUALITY}=5).

\begin{table*}[]
\centering
\caption{List of photometry available as ancillary data for the X-ray sources. The number of detections corresponds to the sample of 888 galaxies initially selected as normal galaxies.}
\label{table:1}
\begin{tabular}{c c c c c c}
\hline
\textbf{Survey} & \textbf{Band} & \boldmath{$\lambda$} \textbf{(µm)} & \textbf{Depth (AB mag)} & \textbf{Number of detections} & \textbf{Reference} \\ \hline
GALEX  & FUV  & 0.15 & 19.9 & 248 & \cite{Bianchi2014} \\ 
         & NUV & 0.23 & 20.8 & 323 &\\ 
\hline
HSC  & g & 0.48 & 26.8 & 830 & \cite{Aihara2018} \\ 
     & r & 0.62 & 26.4 & 573 &\\ 
     & i & 0.77 & 26.4 & 356 &\\ 
     & z & 0.91 & 25.5 & 834 &\\ 
     & y & 0.98 & 24.7 & 835 &\\
     \hline
KiDS/VIKING & u & 0.35 & 24.2 & 576 & \cite{Kuijken2019}\\ 
            & g & 0.48 & 25.1 & 576 & \\ 
            & r & 0.62 & 24.9 & 576 & \\
            & i & 0.76 & 23.7 & 575 & \\
            & J & 1.25 & 21.8 &  573 & \cite{Edge2013} \\
            & H & 1.64 & 21.1 &  573 & \\
            & K & 2.14 & 21.2 &  573 & \\
\hline
LS8 & g & 0.48 & 24.0 &  888 & \cite{Dey2019}\\
    & r & 0.62 & 23.4 &  888 &\\
    & z & 0.91 & 22.5 &  888 &\\
\hline
VISTA/VHS & Ks & 2.15 & 19.8 &  251 & \cite{McMahon2013}\\
\hline
WISE & W1 & 3.35 & 21.0 &  888 & 
\cite{Meisner2019}\\
     & W2 & 4.60 & 20.1 & 888 &\\
     & W3 & 11.56 & 16.7 & 831 &\\
     & W4 & 22.08 & 14.5 & 712 &\\
\hline
\end{tabular}
\end{table*}

\section{Methodology}
\label{Sec:Methodology}

\subsection{SED fitting}
\begin{figure}[]
    \centering
    \includegraphics[width=1\hsize]{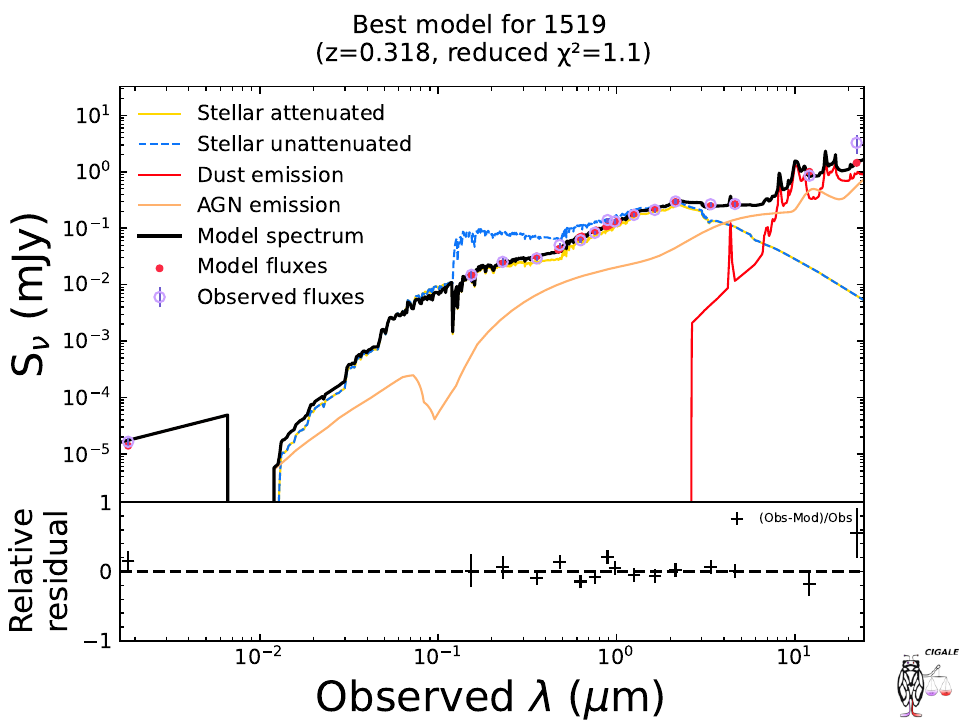}
    
    \caption{Exemplary SED fitted for an object at $z=0.32$, adopting the procedure described in Section~\ref{Sec:Methodology}.  The bottom panel shows the residuals of the fit. }
    \label{fig:SED}
\end{figure}

The SED fitting was performed with the Code Investigating GALaxy emission\footnote{\url{https://cigale.lam.fr}} (CIGALE, \citealt{Noll2009}, \citealt{Boquien2019}). Here, we provide a brief summary of the tool and refer to \cite{Boquien2019} for a detailed description. CIGALE is a Bayesian SED fitting code designed to estimate the  physical properties of the galaxy (e.g., SFR, $M_{star}$). It models the emission spectra of the stellar component and combines them with dust attenuation and emission. The latest version of CIGALE also extends to the X-ray domain, modeling the X-ray emission of the AGN, XRBs, and hot gas components of the galaxy \citep{Yang2021}. In the fitting process, CIGALE preserves the energy balance considering the energy emitted by young massive stars, which is partially absorbed by the dust grains and re-emitted in the MIR and far-IR (FIR).
In this work, the SEDs are built as  the superposition of six modeled components: star formation history (SFH), single stellar population (SSP), dust attenuation, and dust emission, as well as AGN and X-ray emission.  Table~\ref{table:parameters} shows the main input parameters used in the SED fitting process. Figure~\ref{fig:SED} shows an example SED fitted with the adopted procedure. 

\subsubsection{Star formation history and SSP}
As shown in previous works (\citealt{Ciesla2015}, \citealt{Malek2018}, \citealt{Riccio2021}), an SFH that models the bulk of the stellar population with the addition of a recent burst of star formation 
was found to be the best choice to fit statistical samples of galaxies and to reproduce the SFR estimated with other methods (e.g. UV+IR, \citealt{Buat2019}). 
This kind of SFH takes the form of a delayed exponential plus an exponential burst:

\begin{equation}
  \rm{  SFR(t) = t \times e^{(-t/\tau_{main})}/\tau_{main}^2 + te^{-(t-t_{burst})/\tau_{burst}} },
\end{equation}
where $\tau_{main}$ and $\tau_{burst}$ are the e-folding time of the main stellar and the late starburst populations, respectively. 

After calculating the SFH, the following step involves computing the intrinsic stellar spectrum. This requires us to choose a library of single stellar populations. In this work, we adopted the SSP model by \cite{Bruzual2003} with the initial mass function given by \cite{Chabrier2003}. We set the metallicity of the model to $Z=0.02$. The spectrum of the composite stellar populations was calculated by computing the dot product of the SFH with the grid containing the evolution of the spectrum of an SSP with steps of 1 Myr.

\subsubsection{Dust attenuation and emission}

At this stage, the stellar populations are still dust-free. 
As dust attenuation law, we use the \cite{Calzetti2000} law extended with \cite{Leitherer2002} curve between the Lyman break and 150 nm.

To model the dust emission, we employed the \cite{Dale2014} model, based on a sample of SFGs presented in \cite{Dale2002}. 
In their latest update, these authors improved the PAH emission and introduced an optional AGN component. The star-forming component is described with a single parameter $\alpha$, which is defined as d$M_d(U) \propto U^{\alpha}U$d$U$, where $M_d$ is the dust mass and U represents the intensity of the radiation field. The $\alpha$ parameter is closely related to the 60-100 $\mu$m color. The
main strength of this model is its simplicity, with only one parameter that is straightforward to interpret based on the observations. 

\subsubsection{AGN and X-ray emission}
\label{sec:agn_SED}
As shown by \cite{Liu2022}, the eFEDS sample is mainly composed of AGN, especially at high redshifts (\citealt{Mountrichas2022}). We derived the AGN contribution to the UV-to-IR emission of the galaxy using the  \textsc{SKIRTOR} model (\citealt{Stalevski2012}, \citealt{Stalevski2016}). The model considers the primary source of emission of the AGN, the accretion disc, surrounded by an optically and geometrically thick dusty torus. The model allows the user to set several parameters for the geometry of the torus, the extinction and emissivity of the polar dust, its temperature, and so on.

In our analysis, we defined a parameter called $f_{0.25 \mu m}$. This parameter -- as opposed to AGN$_{fraction}$ defined in default by CIGALE as the ratio between the AGN luminosity to the total dust luminosity (between 1-1000 $\mu m $) -- describes the ratio of the AGN UV luminosity at 0.25 $\mu$m to the total UV luminosity at that wavelength. We explain the reason for using the UV estimate, instead of the default parameter in the IR, in Section~\ref{sec:agn_sub}. This parameter can be used to discriminate AGN systems from normal galaxies. 
We sample the  $f_{0.25 \mu m}$ from 0 to 0.7 to consider the strong contribution from nuclear sources in our sample. Allowing the code to consider high nuclear non-stellar contributions gives us the ability to use this parameter to discriminate against AGN systems (see Section \ref{sec:agn_sub}).

The X-ray component is modeled as a superposition of XRBs, hot gas, and AGN.
The HMXBs and LMXBs emission is modeled using the predictions from theoretical XRB population-synthesis models found by \cite{Fragos2013}. However, these relations represent an approximation of the overall population of galaxies and single galaxies can largely scatter around it. For this reason, the \textit{xray} module of CIGALE includes two free parameters, $\delta_{HMXB}$ and $\delta_{LMXB}$ to account for the scatter from the scaling relations. We ran the SED fitting setting $\delta_{HMXB}$ and $\delta_{LMXB}$ to 0 and discuss the possible scatter in the results. The AGN emission was modeled using the $\alpha_{ox}-L_{\nu,2500}$ relation from \cite{Just2007}, where $L_{\nu,2500}$ is the intrinsic disk emission at 2500 $\AA$ at a viewing angle of $30^{\circ}$ and $\alpha_{ox}$ is the AGN SED slope connecting $L_{\nu,2500}$ and $L_{\nu,2keV}$. To consider possible intrinsic X-ray anisotropy, so that an AGN viewed at type 2 angles will have lower fluxes than viewed at type 1 angles, the AGN emission was modelled as a second-order polynomial function of the cosine of the viewing angle (e.g., \citealt{Netzer1987}). More details about the module can be found in \cite{Guang2020, Yang2021}.

The full set of parameters employed in the SED fitting process is shown in Table~\ref{table:parameters}. 
The quality of the fit is expressed by the best $\chi^2$ (and a reduced best $\chi^2$ defined as $\chi^2_r=\chi^2/(N-1)$, with N as the number of data points).  
The minimum value of $\chi^2_r$ corresponds to the best model selected from the grid of all possible computed models from the input parameters.
After the fit, we removed 164 galaxies with $\chi_r^2>10$ from the initial sample of 888 galaxies. Henceforth, we refer to the remaining  724 sources as the final sample. In Table \ref{table:selection}, we summarize sequential steps of the selection criteria, with the respective number of sources selected from the initial sample.

\begin{table*}
\small
\caption{Input parameters for the code CIGALE.}   
\label{table:parameters}      
\centering                          
\renewcommand{\arraystretch}{1.1} 
\begin{tabular}{l| l}        
\hline                
Parameters & Values \\    

\hline
\multicolumn{2}{c}{Star formation history:} \\\hline\hline

       \textit{ Delayed star formation history + additional burst}\\ 
\hline
   e-folding time of the main stellar population model (Myr) & 1000, 3000, 5000 \\
   e-folding time of the late starburst population model (Myr) & 50.0, 100\\
   Mass fraction of the late burst population & 0.0,0.005,0.015, 0.02, 0.05, 0.1, 0.15, 0.20 \\
   Age (Myr) & 8000, 9000, 10000, 11000, 12000\\
   Age of the late burst (Myr) & 100, 150\\

\hline                        

\hline                                   
\hline
\multicolumn{2}{c}{ Single stellar population  \cite{Bruzual2003}} \\\hline\hline

    Initial mass function & \cite{Chabrier2003}\\
    Metallicities (solar metallicity) & $0.02$\\
    Age of the separation between the young and the old star population (Myr) & $10$\\
\hline\hline
\multicolumn{2}{c}{ Dust attenuation law  \cite{Calzetti2000}} \\\hline\hline

    $E(B-V)l$: color excess of the nebular lines &  0.1, 0.3, 0.5, 0.7, 0.9\\
    $E(B-V)f$: reduction factor to compute the E(B-V) for the stellar continuum attenuation&  0.3, 0.5, 0.8, 1\\
\hline\hline
\multicolumn{2}{c}{Dust emission: \cite{Dale2014}}\\ 
\hline\hline

    Fraction of AGN & 0\\
    $\alpha$ slope  & 2.0 \\
    
\hline\hline
\multicolumn{2}{c}{AGN (UV-to-IR): \cite{Stalevski2016}}  \\ 
\hline\hline
Inclination, i.e., viewing angle (\textit{i}) & 30, 70\\

AGN contribution to UV luminosity ( $f_{0.25 \mu m}$) & 0.0, \textbf{0.001}, \textbf{0.0025}, \textbf{0.005}, \textbf{0.0075}, \textbf{0.01}, \\
                          & \textbf{0.025}, \textbf{0.05}, \textbf{0.075}, 0.1, 0.2, 0.3, 0.4,  \\
                         & 0.5, 0.6, 0.7\\
                         
Polar-dust color excess (E(B-V)) & 0, 0.2, 0.4\\
 & \\

\hline\hline
\multicolumn{2}{c}{X-ray emission: \cite{Guang2020, Yang2021}}\\ 
\hline\hline

Photon index  of the AGN intrinsic X-ray spectrum (\textit{gam}) & 1.8\\
    
Power-law slope connecting $L_\nu$ at rest-frame 2500 $\AA$ and 2 keV ($\alpha_{ox}$) & -1.9,  -1.7,  -1.5,  -1.3,  -1.1, -0.9\\

Maximum allowed deviation of $\alpha_{ox}$ from the empirical $\alpha_{ox}-L_\nu$(2500 $\AA$) & 0.2\\
    
Deviation from the expected LMXB scaling relation ($\delta_{lmxb}$) & 0.0\\

Deviation from the expected HMXB scaling relation ($\delta_{hmxb}$) & 0.0\\
\hline

\end{tabular}
\vspace{1mm}
\scriptsize{
 \\{\it \textbf{Notes}:} The input values used for better sampling the $f_{0.25\mu m}$ are in boldface type (see Section~\ref{sec:agn_sub}). }

\end{table*}

\begin{table*}[]
\centering
\caption{Sample selection discussed in Sections~\ref{Sec:Data} and~\ref{Sec:Methodology}}. 
\label{table:selection}
\begin{tabular}{c c c}
\hline
\textbf{Selection criteria} & \textbf{Number of selected sources} & \textbf{\% of the initial sample} \\ \hline
\textsc{CTP\_QUALITY}>2 & 22\,256  & 81\%\\
\textsc{SECURE EXTRAGALACTIC} & 5\,100  & 19\%\\
\textsc{CTP\_REDSHIFT\_GRADE} $\geq 3$ & 5\,100  & 19\%\\
$z<0.35$ & 888  & 3.2\%\\
W1 and W2 $S/N \geq 2$ & 888  & 3.2\%\\
$\chi_r^2<10$$^1$ & 724  & 2.6\%\\
\hline
\footnotesize{
 {\it Notes:} $^1$Selection based on the SED fitting. }
\end{tabular}
\end{table*}

\subsection{Identification of AGN systems}
\label{sec:AGN}

In order to study the properties of galaxies for which the X-ray emission is dominated by XRBs we need to reveal the presence of nuclear non-stellar emission. Given the presence of both nuclear and star-formation emissions, we used a combination of multi-wavelength techniques to identify the AGNs. We estimated the rest-frame X-ray luminosity in the 0.2-2.3 band using the formula:
{\footnotesize
\begin{equation}
  \rm{ log(L_{0.2-2.3}) = log(f_{0.2-2.3})+2log(Dls)+log(4\pi)-log(E_{cor})+log(K_{cor})},
\end{equation}}where $E_{cor}$ and $K_{cor}$  are corrections for the energy range and redshift respectively, \textit{Dls} is the luminosity distance, and $f_{0.2-2.3}$ is the flux detected in the 0.2-2.3 keV band. The assumed photon index is $\Gamma=1.8$, indicated to reproduce emission from HMXBs (\citealt{Lehmer2016}).
We classify a source as AGN if it  satisfies at least one of the following criteria:

\begin{itemize}
    \item X-ray luminosity of $L_{0.2-2.3} \geq 3\times10^{42}$ erg/s;
    
    \item X-ray-to-optical flux ratio of $log(f_X/f_r)>-1$ (where $f_X$ is the flux detected in the 0.2-2.3 keV range, and $f_r$ is the flux observed in the $r$ band);
    
    \item X-ray-to-NIR flux ratio of $log(f_X/f_{K})>-1.2$. 
\end{itemize}   

%
The first three criteria are described in section 4.4 of \cite{Xue2011}.
The above selection criteria may still not identify highly obscured AGN. For this reason, in making use of the MIR observations by WISE, we selected AGN sources following the color selection criteria presented in \cite{Assef2013}. This method selects especially obscured AGN and has the advantage of making no use of the WISE4 band, which is often affected by low S/N. We select as AGN only objects having a 90\% selection reliability. This criterion is particularly sensitive to bright AGNs that also outshine the host galaxy in the WISE bands, but it is less sensitive for obscured AGNs for which the emission is comparable with the host galaxy's. For this reason, although the expected reliability is $\geq 90\%$, the selection completeness drops from 90\% at $W2 \sim 14$ mag to 10\% at  $W2 \sim 16$ mag (see Fig. 4 of \citealt{Assef2013}).

\begin{figure*}[]
    \centering
    \includegraphics[width=0.5\hsize]{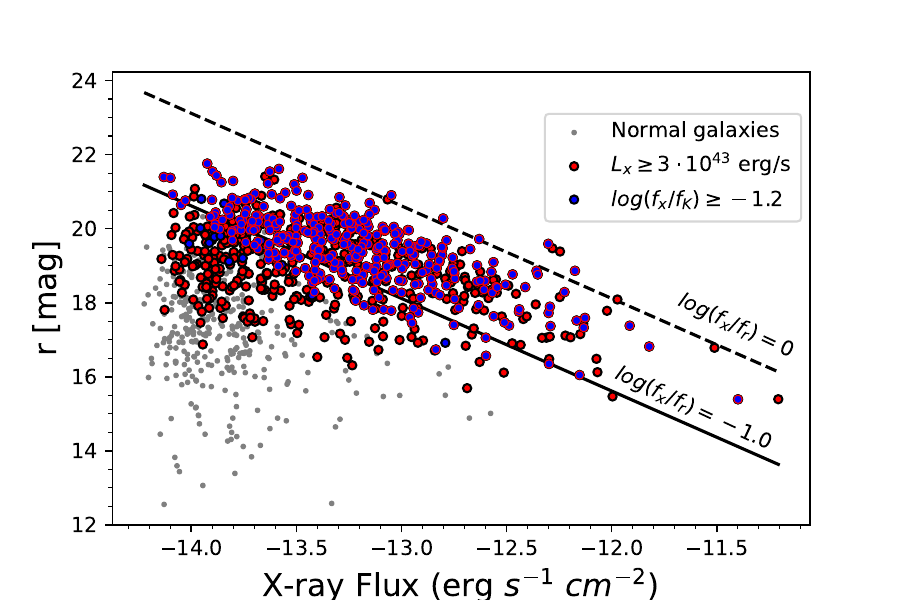}%
    \includegraphics[width=0.5\hsize]{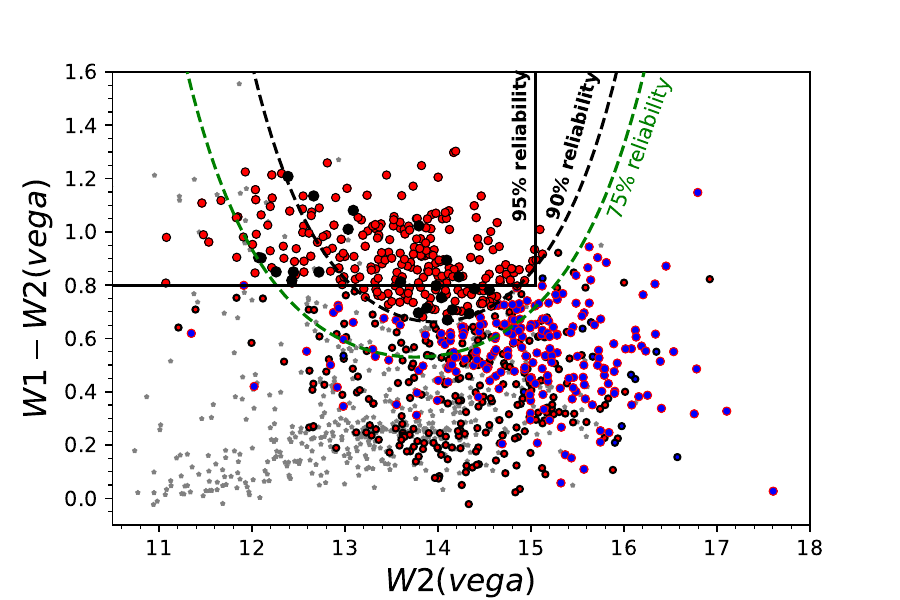} 
    \caption{Optical r-band AB magnitude vs. the X-ray flux for the final selected sample, shown on the left. Diagonal lines indicate a constant flux ratio between the r-band and X-ray.  WISE1-WISE2 color vs. WISE2-band VEGA magnitude, shown on the right. The lines represent the MIR WISE selection criteria, 95\% reliability (solid line), 90\% reliability (dashed black line), and 75\% reliability (dashed green), respectively. Black circles represent the WISE-selected AGNs in this work. For all panels, the grey circles represent the normal galaxies, the red circles the X-ray luminosity selected AGN, blue circles the K band selected AGNs.   }
    \label{fig:AGN_selection}
\end{figure*}

Sources that do not meet any of these criteria are classified as "normal galaxies." We identified 405 AGN (55\% of the final sample) and 319 normal galaxies (44\% of the final sample). Figure~\ref{fig:AGN_selection} shows the X-ray/optical flux selection criteria in the r band (left panel) and the MIR selection criteria adopting WISE photometry. We notice that the majority of the AGNs are selected by X-ray luminosity, though a consistent number of obscured AGNs is selected with WISE selection criteria.  

\begin{figure}[]
    \centering
    \includegraphics[width=\hsize]{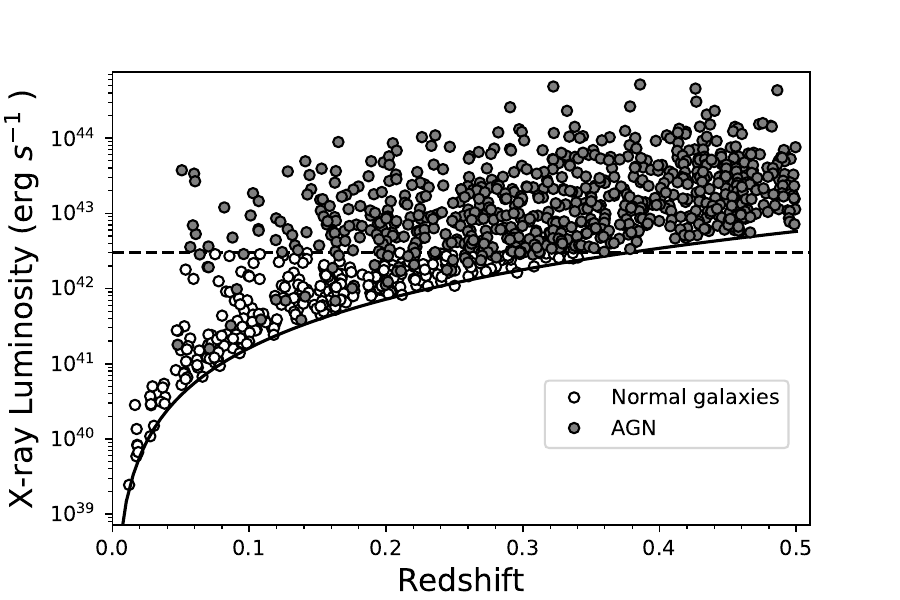}%
    \caption{X-ray luminosity vs redshift for the selected final sample. The dashed black line indicates $L_x=3 \times 10^{42}$ erg/s. The solid black line represent the eFEDS sensitivity limit of $f_{limit}=6.5 \times 10^{-15}$ erg s$^{-1}$cm$^{-2}$ \cite{Brunner2022}. Grey and white circles represent AGN and normal galaxies respectively for the final selected sample.}
    \label{fig:redshift}
\end{figure}

The redshift distribution of AGNs and normal galaxies is shown in Fig.~\ref{fig:redshift}. We notice that above redshift $\sim0.35$, we do not detect any normal galaxy due to the sensitivity limit of the instrument. 

\section{SFR estimates and reliability check}
\label{sec:reliability_check}

\subsection{Mock analysis}

To ensure the reliability of the computed SFR, a mock catalog can be created using an option in CIGALE, which employs the best-fit model from the SED fitting to build an artificial object for each galaxy with known physical parameters. The process is described in detail in  \cite{Giovannoli2011} and \cite{LoFaro2017}.

The physical properties are evaluated using a Bayesian method. This
is done through a likelihood estimation. Each model in
the grid of models built from the starting input parameters will
have an associated likelihood taken as $\propto$ exp($-\chi^2/2$). This value is used as the weight to estimate the physical parameters as the likelihood-weighted mean of the physical parameters attributed to each model, while the related uncertainties are estimated as likelihood–weighted standard deviations of the
physical parameters (see Section 4.3 of \citealt{Boquien2019}).

\begin{figure}[]
    \centering
    \includegraphics[width=0.8\hsize]{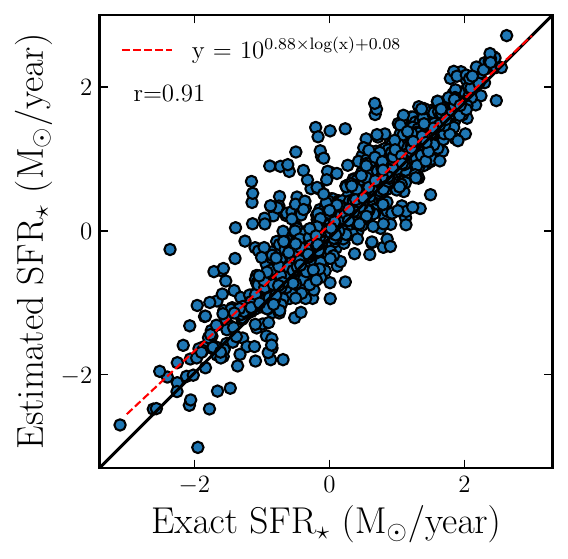}%
    \caption{Comparison between the true value of the SFR provided by the best-fit model for the mock catalog (x-axis) and the value estimated by the code (y-axis). The Pearson product-moment correlation coefficient is given as the ‘r’ value. The black line corresponds to the 1:1 relation, while the red dashed line is the regression line with the equation given in the legend.}
    \label{fig:reliability_SFR}
\end{figure}

In Figure \ref{fig:reliability_SFR}, we compare the output SFR of the mock catalog with the best values estimated by the code. The Pearson product-moment correlation coefficient (r) is used as a measure of the reliability of the obtained properties. We find a slight overestimation of the estimated values at low SFR, but they are still statistically consistent with the exact value.

\subsection{SFR estimates with FIR data}

As shown in previous works (i.e., \citealt{Buat2019}, \citealt{Riccio2021}), the lack of IR observations can lead to over- or underestimations of the SFR when broadband SED fitting methods are used. In particular, when FIR data are missing, it is not possible to constrain with sufficient precision the peak of the dust emission, making the estimate of the total dust luminosity incomplete. This could lead to inaccurate implementation of the energy balance and, finally, to over- or underestimations of the SFR. Even though \cite{Riccio2021} showed that a combination of optical and MIR data is enough to reliably estimate the SFR at these redshifts, to be as conservative as possible, we validated our SFR estimates (when possible) using other indicators.  

To validate  SFRs obtained from the broad-band SED fitting, we acquired data in the FIR wavelength to improve the estimation of the dust luminosity.  For this reason, we cross-matched our sample with observations performed by for the \textit{Herschel Extragalactic Legacy Project} (HELP, \citealt{Shirley2021}) survey in the GAMA09 field, which overlaps with the eROSITA data. Herschel was equipped with two imaging instruments, the Photodetector Array Camera and Spectrometer (PACS; \citealt{Poglitsch2010}), which observed the FIR at 100 and 160 $\mu$m, and the Spectral and Photometric Imaging Receiver (SPIRE; \citealt{Griffin2010}), which covered the 250, 350, and 500 $\mu$m wavelength ranges. In the GAMA09 field, we only found  detections from the SPIRE instrument.  We identified 53 matches to our sample, adopting a $1\arcsec$ matching radius using coordinates from the optical observations. The cross-match was done only on galaxies that we flagged as "normal galaxies" with the AGN selection discussed above (in Section~\ref{sec:AGN}). Henceforth, we refer to the objects with HELP counterparts as the eROSITA$_{G9}$ sample. To estimate the number of false matches, we shifted the FoV of the eROSITA sample of $10\arcsec$ in all directions, each time performing the cross-match again. We find one possible false match between the two samples. We then perform the SED fitting on the eROSITA$_{G9}$ sample using UV, optical, and WISE data from the initial sample, plus SPIRE data from HELP. We remove possibly failed fit cutting sources with $\chi^2_r>10$. This cut removes five galaxies from  eROSITA$_{G9}$  sample. 

\begin{figure}
    \centering
    \includegraphics[width=0.9\hsize]{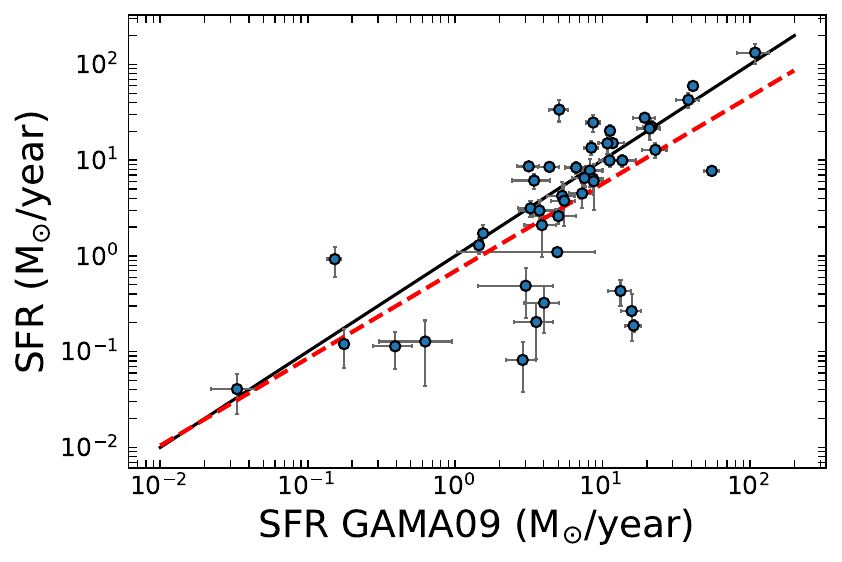}
    \includegraphics[width=0.9\hsize]{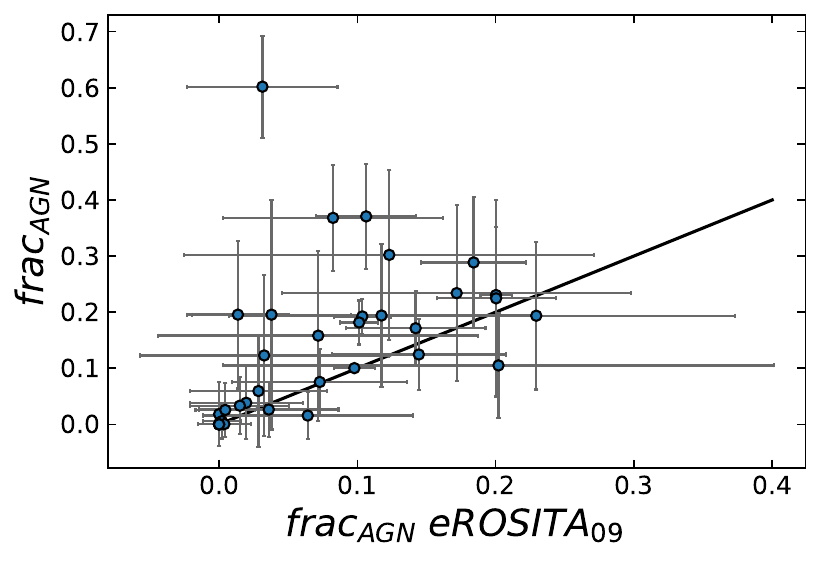}
    \caption{ Comparison between the SFR evaluated with data up to WISE (y-axis) and SFR estimated adding SPIRE FIR data (x-axis), shown at the top. The black solid line corresponds to the 1:1 relation, red dashed line to the linear fit of the data. Comparison between $frac_{AGN}$ parameter estimated by the two runs, shown at the bottom. The solid line corresponds to the 1:1 relation.}
    \label{fig:SFR_gama}
\end{figure}

Figure~\ref{fig:SFR_gama} shows the results obtained with SED fitting using data up to WISE detections, with the one obtained for the eROSITA$_{G9}$ sample. The top panel shows overall comparable results for the SFR, with a slight underestimation at $SFR>1$. Furthermore, the employ of MIR observation without FIR detections can lead to a wrong differentiation between dust emission due to the AGN activity and due to star formation. Indeed, comparing the  $f_{0.25 \mu m}$ parameter between the two runs, we find higher values when only data up to the MIR wavelengths are used (bottom panel, Fig.~\ref{fig:SFR_gama}). The change in the $f_{0.25 \mu m}$ value while constraining the peak of the dust emission using SPIRE data, can be explained by two scenarios: 1) the code correctly attributes to the star formation activity part of the MIR emission assigned previously as an AGN contribution, resulting in a lower  $f_{0.25 \mu m}$ or  2) Herschel observations globally increase the IR luminosity, and thus the SFR, at a fixed AGN luminosity, resulting in lower $f_{0.25 \mu m}$. However, in checking the dust luminosity estimated with the inclusion of Herschel with the one calculated using up to MIR, we find comparable results similar to what was found for the SFR. This result supports the first scenario. As a result of the comparison, only for the 48 objects having Herschel counterparts, we decided to update the values of the SFR,  $f_{0.25 \mu m}$, and the other physical parameters with those obtained from the SED fitting of the eROSITA$_{G9}$ sample. In doing so, we do not expect to introduce any systematics in the SFR, as both estimates are comparable and trace the same timescales of star formation.

\subsection{Spectral counterpart and BPT diagram}
\label{sec:BPT}
As discussed in Section~\ref{sec:Introduction}, the emission from HMXBs directly traces the young stellar population. This makes the $H_\alpha$ SFR indicator the best to study the Lx-SFR relation, as it traces the emission from young stellar populations with the resolution of a few million years. For this reason, to further check the reliability of our SFR estimates, we cross-matched our sample with the MPA/JHU catalog based on the \textit{Sloan Digital Sky Survey} DR7 release (\citealt{Abazajian2009}), which provides images, imaging catalogs, spectra, and redshifts. We are interested in the spectral data of the catalog, especially the $H_\alpha$ line, whose specifications are explained in detail in \cite{Tremonti}. The cross-match was performed in the same way as the eROSITA$_{G9}$ sample. We identify 106 matches to our sample adopting a $1\arcsec$ matching radius. Again, the cross-match is restricted only to galaxies flagged as "normal galaxies." We identified one possible false match between the two samples. To avoid biases in the SFR estimates, we choose sources having $S/N>3$ for the $H_\alpha$ line, leaving us with 34 counterparts.

For the 34 galaxies, we corrected 
$H_\alpha$ and $H_\beta$ emission lines using Balmer decrement 
assuming \cite{Calzetti2000} attenuation law with $Rv=3.1$ (see Eq.~6 in \citealt{Yuan2018}).  After correcting for the attenuation, we employ \cite{Kennicutt98} relation to estimate the SFR from $H_{\alpha}$ line.

\begin{figure}
    \centering
    \includegraphics[width=0.9\hsize]{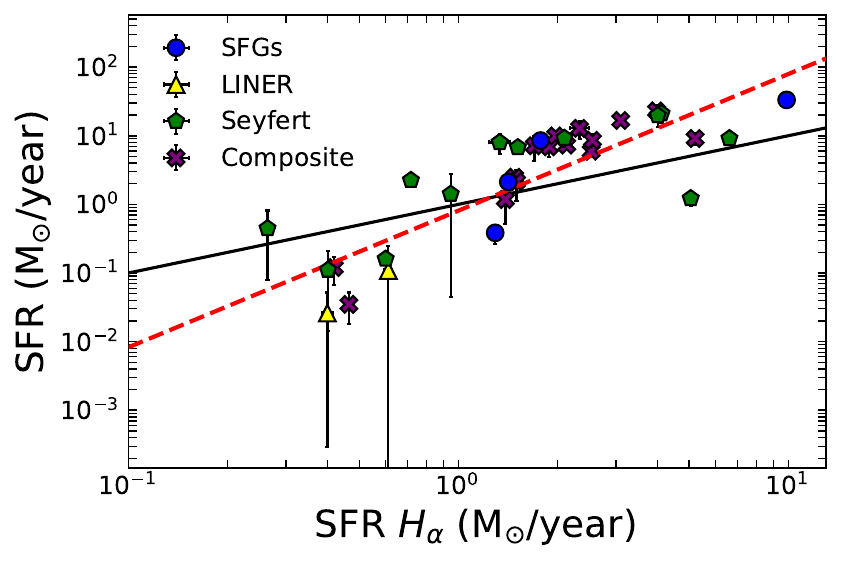}
    \includegraphics[width=0.9\hsize]{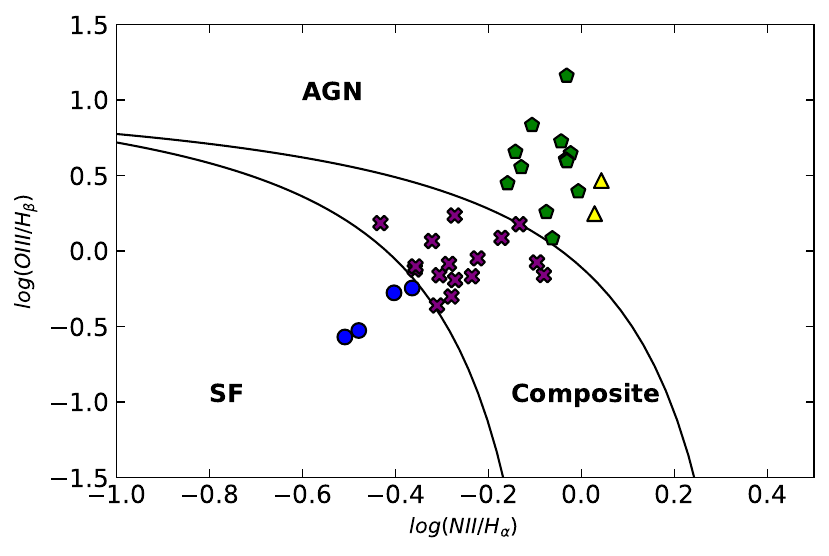}
    \caption{Comparison between the SFR evaluated with the SED fitting up to WISE (y-axis) and SFR estimated using $H_\alpha$ lines (x-axis), shown at the top. The black solid line corresponds to the 1:1 relation, red dashed line to the linear fit of the data. Bottom panel shows one of three BPT diagrams used to select normal star-forming galaxies for the sources having spectral counterparts. The solid line delimits the division between SFG, AGNs, and composites. For both panels, the blue dots represent the SFGs, yellow triangles: LINERs, green pentagons: Seyferts, and purple crosses: composite galaxies
  }
    \label{fig:sfr_alpha}
\end{figure}

The top panel of Fig.~\ref{fig:sfr_alpha} shows the comparison between derived from ${H\alpha}$ (hereafter, $\rm SFR_{H\alpha}$) and the one from the SED fitting up to WISE detection. This comparison shows a consistent difference between the two values of the SFR, with the SED fitting underestimating the SFR below 2 $M_{\odot}$yr$^{-1}$, and overestimating above this value. To explain this trend, we examine the Baldwin-Phillips-Terlevich diagram (BPT, \citealt{Baldwin1981}) to further inspect the presence of nuclear activity due to SMBH accretion. 
The bottom panel of Fig.~\ref{fig:sfr_alpha} shows one of the  BPT diagrams used for our analysis of the 34 sources flagged as "normal galaxies" with SDSS spectra. Galaxies were classified using the optical emission line ratios $\log([N_{II}]\lambda6584/H_\alpha)$, $\log([S_{II}]\lambda6717,6731/H_\alpha$, $\log([O_I]\lambda6300/H_\alpha)$, and $\log([O_{III}]\lambda5007/H_\beta)$, as star-forming galaxies (SFGs),  \textit{Low-ionization nuclear emission-line region}  (LINER), Seyfert, and composite, according to their position in the three BPT diagrams. We find that the majority of the galaxies that scatter from the 1:1 relation are classified as Seyfert or LINER. These galaxies are known to host ionization of the ISM, which can be powered both by star formation or AGN activity (\citealt{Heckman}, \citealt{Terlevitch}). However, most of these galaxies in the nearby Universe appear to have low levels of star formation activity (\citealt{Larkin}, \citealt{Bendo}), and the mid-infrared spectra do not appear similar to the spectra expected from star formation. For these objects, the $H_\alpha$ emission would be attributable to the activity of the SMBH and so would not be appropriate to use it as SFR indicator. For this reason, at the end of the analysis, we decide to update the SFR estimates with the one obtained with $H_\alpha$ only for the sources classified by the BPT diagram as star-forming and to exclude AGN and composite galaxies from our sample. Also, this investigation suggests that our sample of normal galaxies could be affected by a severe AGN contamination.

At the end of the process, 48 galaxies are updated with the SFR estimated using SPIRE data,  eROSITA$_{G9}$ sample, and four galaxies are updated with $H_\alpha$ estimated values, for a total of 52 sources (7.3\% of the "normal galaxy" sample). We stress that the SDSS subsample and the eROSITA$_{G9}$ sample do not have any common source.

\subsection{Galaxies distribution on the SFR-$M_{*}$ relation}
\label{sec:MS}

\begin{figure*}[]
    \centering
    \includegraphics[width=\hsize]{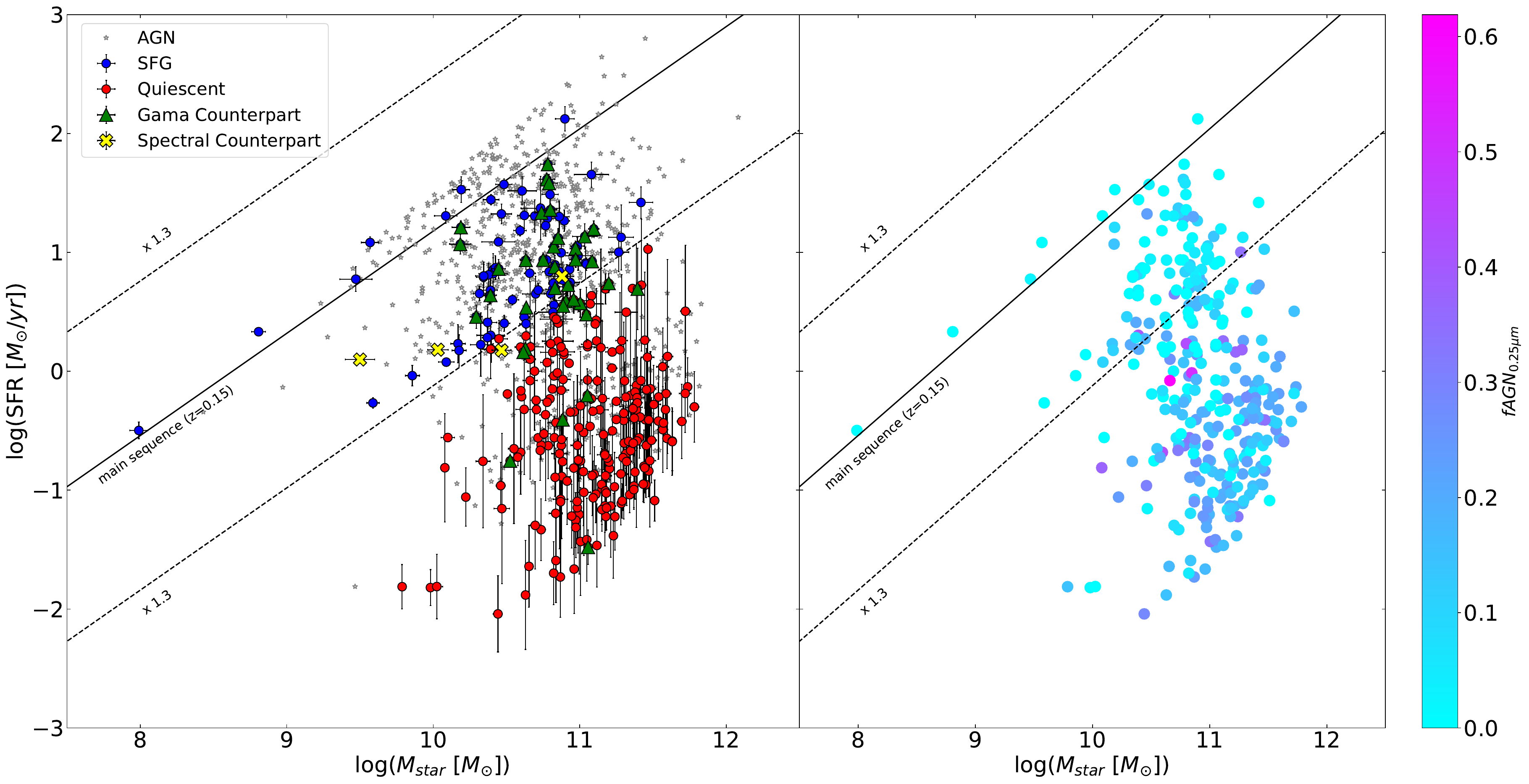}%
    \caption{ 724 galaxies from the final sample shown in the SFR-$M_{*}$ plot (left). Blue points represent the SFGs, red points show the location of the quiescent galaxies, and grey points the AGNs classified as described in section \ref{sec:AGN}.  Right panel displays the same plot showing star-forming and quiescent galaxies color-coded by the $f_{0.25\mu m}$ parameter. The solid black line represents the MS from \cite{Aird2017} at redshift 0.15 (the mean redshift of our sample of normal galaxies). The dashed line  shows the threshold 1.3 dex above and below the MS. }
    \label{fig:MS}
\end{figure*}

It is known that the SFR-$M_{*}$ plot of the galaxies highlights the existence of three different primary populations, according to their efficiency to form stars. Figure \ref{fig:MS} shows the distribution of our sample of galaxies on the SFR-$M_{*}$ plot. We classified them based on their SFRs relative to the evolving star-forming main sequence. We set the threshold between categories as 1.3 dex below (or above) the main sequence (MS) defined by \cite{Aird2017} and given by the equation: 
\begin{equation}
    \log \text{SFR}(z)\ [\mathrm{M}{\odot}\mathrm{yr}^{-1}] = -7.6 + 0.76 \log \frac{M_{\ast}}{M_{\odot}} + 2.95 \log(1+z).
    \label{eq:1}
\end{equation}
Galaxies 1.3 dex below the MS are categorized as quiescent or passive, those above 1.3 dex are classified as starburst, while those lying in between are labeled as SFGs. It is worth noting that the relation presented in  Eq.~\ref{eq:1} is redshift-dependent. To account for this dependence, we use the redshift of each individual object to classify it as passive, normal star-forming,  or starbursting. 

Using Eq.~\ref{eq:1} we find that our final sample of 319 galaxies consists of 98 sources classified as star-forming (30\% of the normal galaxy sample) and 221 as quiescent galaxies (70\% of the normal galaxy sample). 
Our sample does not include any starbursting candidates. 

The right panel of Fig.~\ref{fig:MS} shows the selected normal galaxies color-coded by $f_{0.25\mu m}$ parameter. We find a wide range of $f_{0.25\mu m}$ values. 
We stress that the majority of the sources with high $f_{0.25}\mu m$ values, higher than 0.2, reside in the region populated by the quiescent galaxies. 
This further suggests that our sample of normal galaxies can be strongly contaminated by nuclear activity. We discuss this contamination in Section~\ref{sec:agn_sub}.

\section{Subtraction of other contributions to the total X-ray luminosity}
\label{sec:substraction}

As discussed in section \ref{sec:Introduction}, the X-ray emission of galaxies is the combination of the output of different sources, such as XRBs, hot gas, and AGNs. Different types of galaxies have varying contributions to their X-ray emissions. In normal SFGs, the XRBs typically dominate the total emission at energies $\sim1-10$ keV, with LMXBs associated with the old stellar population in the bulge while HMXBs are linked to younger stellar population concentrated primarily in the disk or in the arms of the spiral. Conversely, as usually undergoes a prolonged period of quenching, elliptical galaxies have only one type of XRB, which is the LMXBs.  While the hard band (2-10 keV) is entirely dominated by XRBs in normal galaxies, the soft band (0.2-2.3 keV) can be significantly contaminated by diffuse intracluster hot gas emission, especially in elliptical galaxies. This contribution was found in the literature to scale with the SFR in SFGs (\citealt{Mineo2012}) and with K-band luminosity in ellipticals (\citealt{Kim2013}, \citealt{Civano2014}). Therefore, it is necessary to adjust the X-ray luminosity by accounting for the various types of galaxy populations.

Coronally active binaries (ABs) and cataclysmic variables (CVs) are additional types of stellar sources that can emit X-rays, hence making a contribution to the total X-ray luminosity of a galaxy. Although their X-ray luminosity was estimated in elliptical galaxies (e.g., \citealt{Pellegrini1994}), their significance is often disregarded due to their relatively weaker luminosities compared to the more luminous LMXBs (see \citealt{Fabbiano2006} for a review). On top of that, even if it is not entirely dominated by AGNs, a nuclear contribution to the total X-ray luminosity can still be relevant. To account for all listed above sources of emission, we performed a further cut of AGN systems, based on the SED fitting and a correction of the X-ray luminosity of our sample, separately for the two populations of galaxies selected in Section~\ref{sec:MS} (quiescent and SF).

\subsection{AGN contamination to the total X-ray luminosity}
\label{sec:agn_sub}

As discussed in Section \ref{sec:BPT}, some of the sources previously classified as "normal galaxies" are nevertheless identified as LINER or Seyfert galaxies by the BPT diagram. This makes clear that the classification carried out in Section \ref{sec:AGN} is not enough to ensure a reliable sample of normal galaxies, and further investigations must be performed.

As described in Section \ref{sec:agn_SED}, the SED fitting process provides us with further information, (independent of the above) to classify AGNs via the parameter $f_{0.25\mu m}$.
A conservative choice often adopted in previous works (\citealt{Malek2018}, \citealt{Padilla2022}, \citealt{Suleiman2022}) defines galaxies with negligible AGN contribution the one having $fAGN\leq0.1-0.2$~, depending on the work. However, this limit employed in the literature referred to the $fAGN$ parameter estimated in the IR range. In fact, as discussed in section \ref{sec:agn_SED}, the $fAGN$ parameter represents the fraction of the emission attributed to the AGN over the total emission of the galaxy in a specific wavelength, which can be set in CIGALE. Thus, as we estimated the parameter in the UV range (at 0.25 $\mu$m), we had no previous literature feedback to identify a limit to safely select AGN systems. Therefore, further investigation must be carried out.
We present the results as a function of $f_{0.25 \mu m}$, estimated at 0.25 $\mu$m, for two primary reasons.  First, our objective is to obtain information on the X-ray emission of these sources, and since CIGALE computes this emission using the $L_{2500 \AA}$, the  $f_{0.25 \mu m}$ is directly linked to the $Lx_{AGN, 0.2-2.3}$. 

Second, probing the  $f_{0.25 \mu m}$ parameter allows us to identify Seyfert~1 galaxies, which are the main type of AGNs expected to contaminate our sample of  normal galaxies. Indeed, obscured AGNs (Seyfert~2) are expected to be well classified with the WISE band selection, with expected reliability of $\sim90\%$ (Section~\ref{sec:AGN}, and  Fig.~\ref{fig:AGN_selection} right panel). Also, highly obscured AGNs for which it is not possible to observe the broad line region (and therefore the  $f_{0.25 \mu m}$ would lead to wrong findings) are expected to be considerably obscured in the X-ray regime.

The basic aim presented in this section is to select a sample of normal galaxies with the parameter $fAGN_{0.2-2.3keV}$ defined as:
\begin{equation}
    \rm{fAGN_{0.2-2.3keV} = L_{AGN, 0.2-2.3keV}/L_{total, 0.2-2.3keV}},
\end{equation} 
which is less than an arbitrary threshold. Unfortunately, the standard version of CIGALE does not directly allow to estimate the $f$ in the X-ray band due to its structure: the \textit{xray} module is added to the SED fitting process after the AGN module when the $f_{0.25 \mu m}$ is already computed. For this reason, we added the possibility to estimate $fAGN_{0.2-2.3keV}$ in the \textit{xray} module as part of the Bayesian evaluation process. 
This parameter will be strictly connected to the  $f_{0.25 \mu m}$ through the $L_{AGN, 0.2-2.3keV}$ luminosity, underlining the importance of a reliable estimate of this parameter.
Therefore, we run the SED fitting process again, only on the sample of normal galaxies, better sampling the  $f_{0.25 \mu m}$ for low values\footnote{The list of the $f_{0.25 \mu m}$ parameters used for this second, refined  SED fitting, is marked with bold values in the Table~\ref{table:parameters} }. We limit the set of parameters to low values as we want to increase the quality of the $fAGN_{0.2-2.3keV}$ estimates for $f_{0.25 \mu m}$ lower than 0.1. With this set of parameters, all the sources previously best-fitted with $f_{0.25 \mu m}>0.1$ will have catastrophic fits and will be removed from the sample. 

\begin{figure}[]
    \centering
    \includegraphics[width=0.9\hsize]{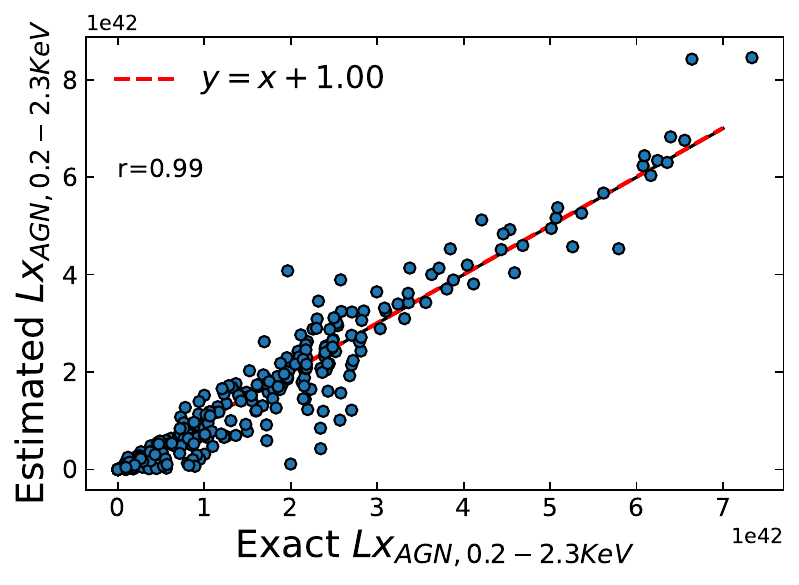}
    \includegraphics[width=0.9\hsize]{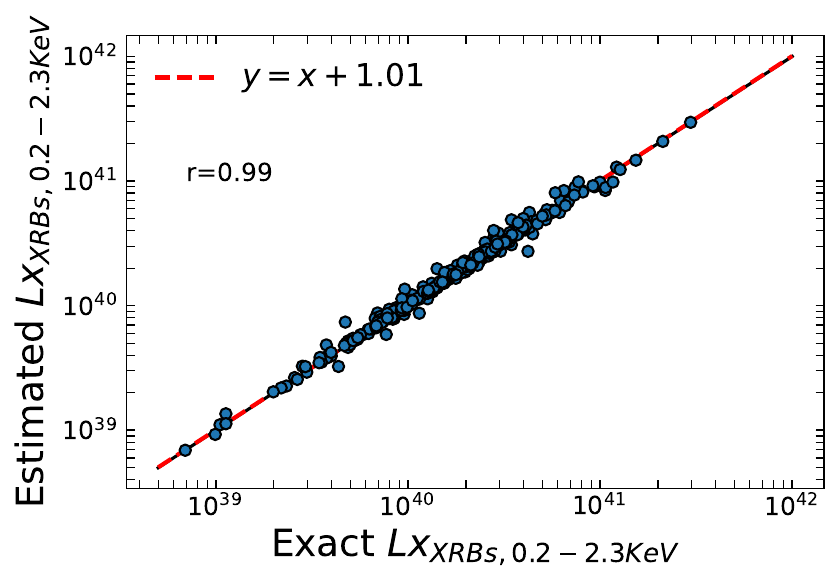}
    \includegraphics[width=0.9\hsize]{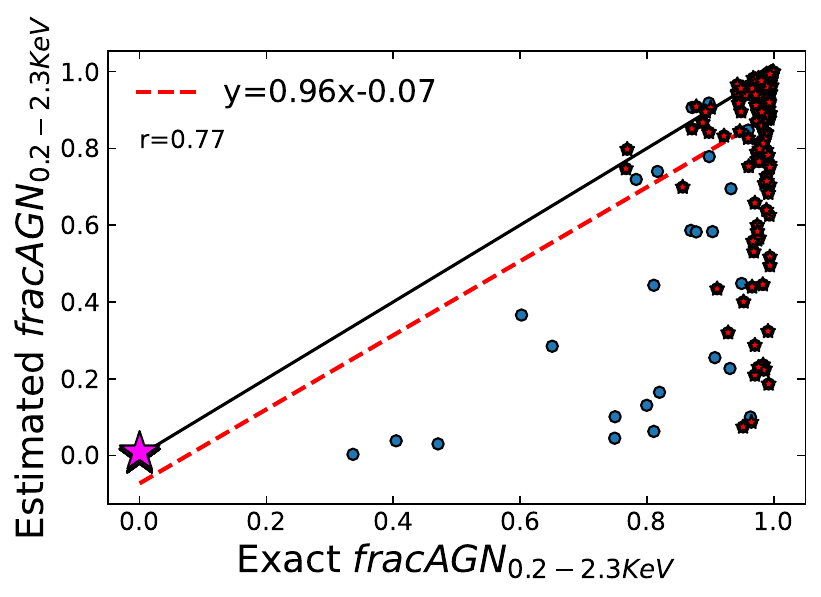}
    \caption{ Mock analysis for the $L_{AGN, 0.2-2.3keV}$ (top panel), $L_{XRB, 0.2-2.3keV}$ (middle panel), and $fAGN_{0.2-2.3keV}$ (bottom panel).  The Pearson product-moment correlation coefficient is given as an ‘r’ value. The black solid line corresponds to the 1:1 relation, while the red dashed line is the regression line with the equation
    given in the legend. The magenta star represents the superposition of 44 sources. The red stars are the sources having $f_{0.25\mu m}>0.1$ from the initial run described in Section~\ref{Sec:Methodology}.   }
    \label{fig:mock_Lx}
\end{figure}

As our goal is to select sources with minimal  $fAGN_{0.2-2.3keV}$ contribution, to study the relation between SFR and the Lx for normal galaxies, the cleaning described above will not affect the final results. Furthermore, the estimates of SFR and $M_{star}$ will not be significantly affected by the change since their estimation depends mostly on the SFH, SSPs and dust-related modules.
In Fig.~\ref{fig:mock_Lx} we show the mock analysis for the computed values of $L_{AGN, 0.2-2.3keV}$, $L_{XRB, 0.2-2.3keV}$ and $fAGN_{0.2-2.3keV}$. 
We notice that the AGNs and XRBs contributions to the X-ray luminosity are statistically well estimated by the SED fitting. Instead, we find a slight difference between the estimated and the exact values for the $fAGN_{0.2-2.3keV}$ parameter, with a Pearson correlation coefficient of $r=0.77$. However, the difference is mainly carried out by sources with estimated $fAGN_{0.2-2.3keV}$ greater than 0.2, which will be removed at the end of the process.
This result can be explained by considering the functioning of the Bayesian method. 
The sources far from the 1:1 relation have considerably higher values of $fAGN_{0.2-2.3keV}$ estimated from the best-fit SED  compared to the one estimated by the Bayesian method. 
For these sources, many models predicting low values of $fAGN_{0.2-2.3keV}$ equally well fit the photometry. Consequently, these models will be associated with a high weight in the weighted estimation of the physical properties, considerably lowering the value of the estimated $fAGN_{0.2-2.3keV}$.
On the other hand, as we expected, the majority of the sources that are clustered on the right part of the plot are associated with a value of  $f_{0.25 \mu m}$ higher than 0.1 in the initial run described in Section~\ref{Sec:Methodology}. Therefore, the sources we are interested in are in the bottom left part of the diagram. We stress that this region of the plot, where the sources have estimated values of $fAGN_{0.2-2.3keV} \sim 0$ (but never exactly 0), consists of 44 galaxies (magenta star in Fig.~\ref{fig:mock_Lx}).

To be conservative, we decided to make our selection of normal galaxies on the basis of those expected to have an AGN contribution to the X-ray luminosity that is less than 10\% ($fAGN_{0.2-2.3keV}<0.1$). With this limit, not only are we able to cut the majority of the outliers of the mock analysis, but we are sure to analyze only the stellar component of the X-ray emission.
In this way, we select 47 sources: 32 SF and 15 quiescent galaxies. Only seven of these sources have an exact value of  ${fAGN_{0.2-2.3keV}}$ larger than 0.1. 
We stress that all the sources below the threshold have estimated $f_{0.25 \mu m} \sim 0$. At the end of the process, for the final sample of 47 galaxies, we expect to have X-ray emission from XRBs + possible hot gas and CVs/ABs components. In the following subsections, we discuss the subtractions of these components.

\subsection{Quiescent galaxies}

 To evaluate the contribution of the hot gas to the total X-ray emission of the passive galaxies, we used the relation between the X-ray emission and K-band luminosity in the form of $L_x \sim L_K^\alpha$, with an exponential slope of $\alpha = 4.5$ (\citealt{Kim2013}, \citealt{Civano2014}).
As the $K$ magnitude, we used the VISTA/VHS $K_S$ band available as ancillary data in the catalog from \cite{Salvato2022}. We calculate the $K$ band luminosity in units of solar luminosities using the equation from \citealt{Civano2014}:
\begin{equation}
  \rm{  L_{K}[L_\odot]=10^{(-K-K_\odot)/2.5} \cdot (1+z)^{\alpha-1} \cdot (D_{Ls}/10)^2},
    \label{Eq:quie_hotgas}
\end{equation}
where $K$ is the $K_S$ magnitude from VISTA/VHS, $z$ is the redshift, $D_{Ls}$ is the luminosity distance in parsec, and $K_\odot = 5.12$ is the absolute AB magnitude of the Sun in K-band. To estimate the luminosity, a spectral shape of the type $f_\nu = \nu^\alpha$ is assumed, where $\alpha=-(J-K)/\log(\nu_J/\nu_{K})$ and $J-K$ is calculated from the magnitudes.

To account for the emission from ABs and CVs we use the relation found in \cite{Boroson2011} for the soft band (0.5-2 keV):

\begin{equation}
   \rm{ L_X [erg/s] =4.4^{+1.5}_{-0.9} \times 10^{27} L_K [L_\odot] }
    \label{Eq:boroson}.
\end{equation}

\subsection{Star-forming galaxies}

For SFGs, it is of crucial importance to isolate the contribution produced by HMXBs to the total X-ray emission of the galaxy, as this component correlates directly with the SFR. For this reason, we estimate the X-ray contribution from LMXBs, hot gas and ABs and CVs. 

To account for the contribution of LMXBs, we employed the relation between $L_{x,LMXBs}$ and $M_*$ found by \cite{Gilfanov2004}. In this work, they study the properties of X-ray binaries in 11 local early and late-type galaxies, finding that, in late-type galaxies, the $L_{x, LMXBs}$ correlates with the $M_*$ as:

\begin{equation}
  \rm{  \frac{L_{x,LMXBs}}{10^{40}} erg/s = \frac{M_*^{0.98} M_\odot}{10^{11} L_\odot}.}
    \label{Eq:LMXBs}
\end{equation}
 
As well as early-type galaxies, SFGs are also known to possess a significant amount of hot ionized gas,  which is the source of X-ray emission and which was found to correlate with their SFR. To account for this component, we make use of the relation found in \cite{Mineo2012}. Using a sample of  nearby late-type galaxies, they found that the X-ray luminosity due to hot gas correlates with the SFR as:
\begin{equation}
   \rm{ L_{0.5-2keV}^{gas} (erg/s)=(8.3 \pm 0.1) \times 10^{38} \cdot SFR (M_\odot yr^{-1})}.
    \label{Eq:hot}
\end{equation}
To determine the combined X-ray emission from AB+CV, we use the same relation as for quiescent galaxies. 
The X-ray luminosities estimated from Eqs.~\ref{Eq:boroson} and~\ref{Eq:hot} are converted from the 0.5-2 keV to the 0.2-2.3 keV band assuming a power law photon index of $\Gamma=1.26$ (\citealt{Boroson2011}) and  $\Gamma=1$ (\citealt{Mewe1986}), respectively.

\begin{figure*}[]
    \centering
    \includegraphics[width=0.45\hsize]{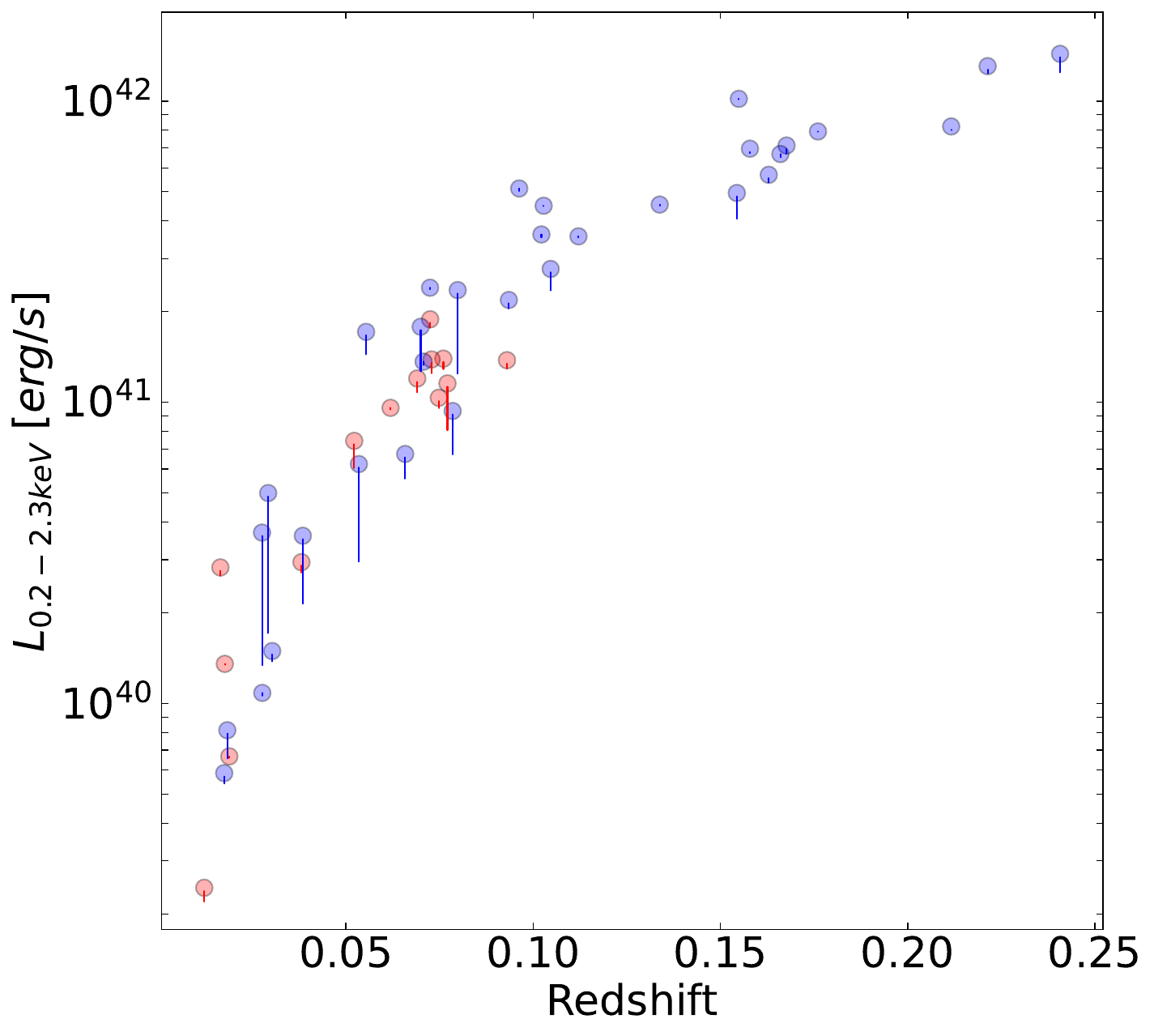}%
    \includegraphics[width=0.420\hsize]{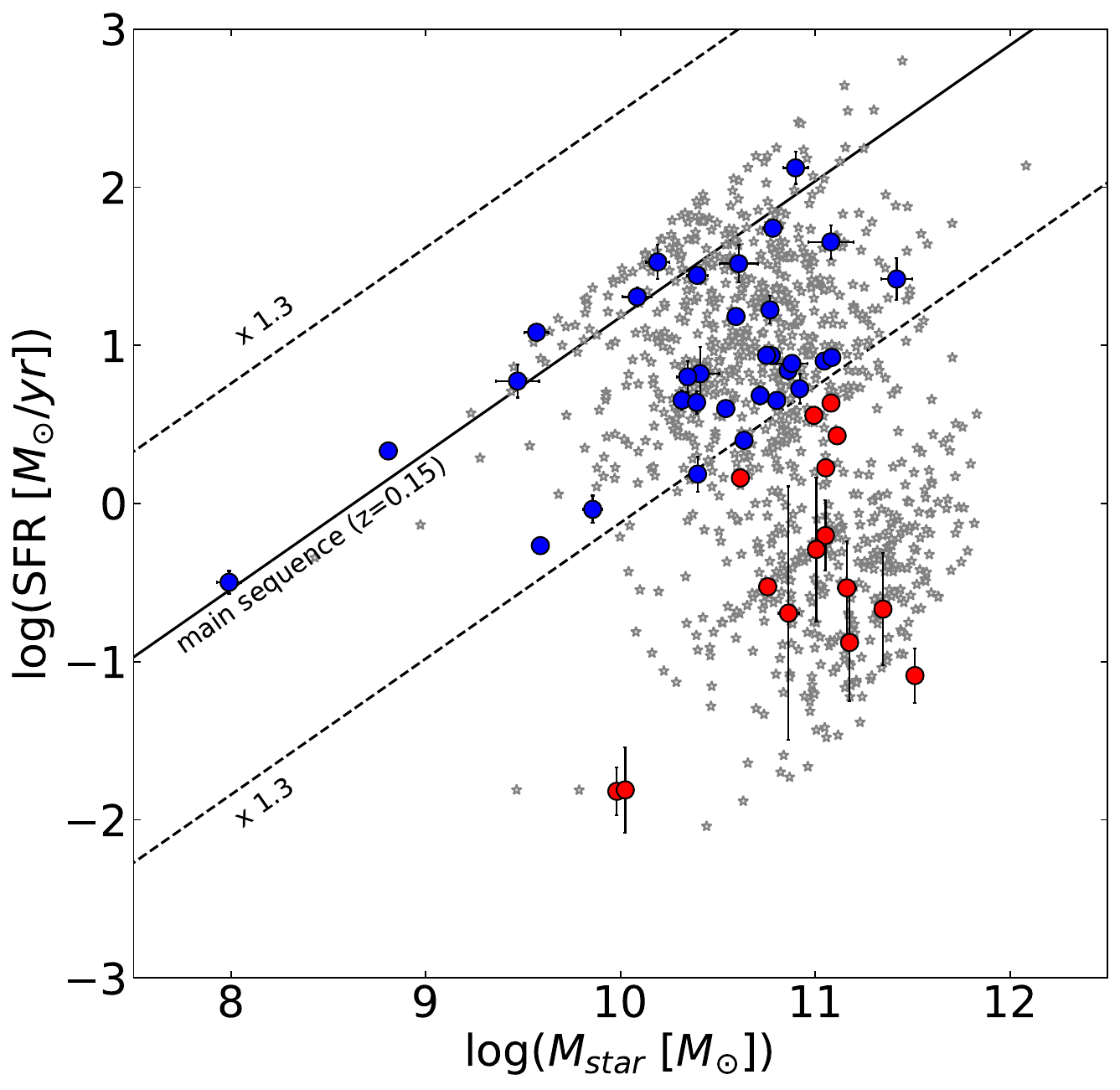}%
    \caption{X-ray luminosity distribution over redshift for our sample of quiescent (red circles) and SFGs (blue circles) in the left panel. The uncorrected values are presented as solid circles while the change in the $L_x$ after the correction is shown by a solid line. Position of the final sample of 47 "normal galaxies" on the MS in the right panel. Quiescent galaxies are represented by red circles, while SFGs as blue circles. AGN systems are shown in grey.}
    \label{fig:quiescent_sub}
\end{figure*}

Figure~\ref{fig:quiescent_sub} shows the redshift distribution of the sample of quiescent and SFGs after the subtractions discussed above. We notice that for SFGs at low redshift ($z=0-0.1$), the subtracted X-ray luminosity attributed to hot gas, LMXBs, and CVs and ABs reaches on average 20\% of the observed X-ray luminosity. This contribution drops to an average of 3\% going to higher redshift. The figure also shows the position of the final sample of 47 "normal galaxies" on the MS. 

\section{Lx-SFR relation}
\label{Sec:Results}
\begin{figure}[]
    \centering
    \includegraphics[width=\hsize]{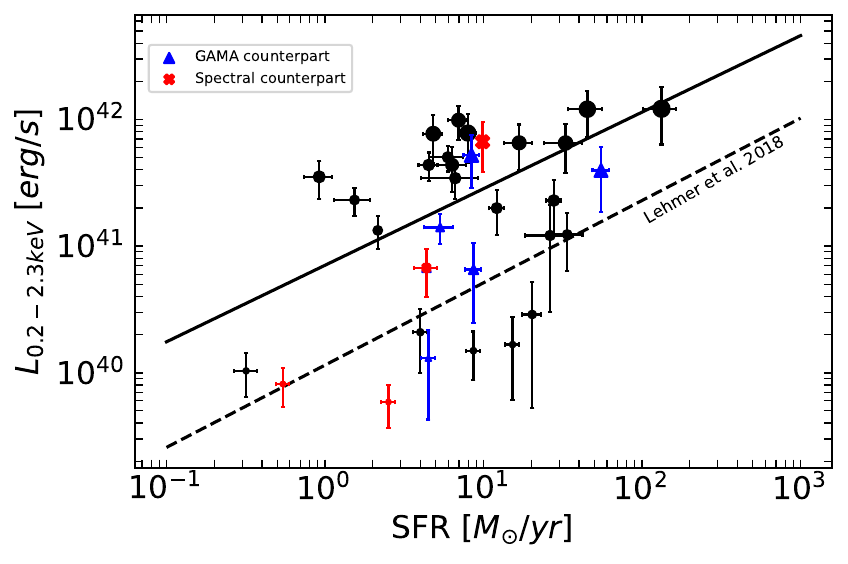}%
    \caption{ X-ray luminosity as a function of the SFR. Black dots represent the full sample of SFGs. Red crosses and blue triangles represent the sources having \textit{Herschel} and SDSS counterparts respectively. The solid black line represents the fit of our sample of data, while \cite{Lehmer2016} scaling relation is represented by the dashed black line. The size of the dots is proportional to the redshift of the sources. }
    \label{fig:Lx_SFR}
\end{figure}

As discussed in Section~\ref{sec:Introduction}, our primary goal is to constrain the connection between X-ray luminosity and star formation activity for the sample of X-ray detected normal galaxies observed in the eFEDS survey. 
This can be achieved by fitting  the empirical scaling relation between X-ray luminosity and SFR and comparing it with those inferred for local and distant galaxies. 
We limit this analysis only to the SFGs, as it was shown to have a strict dependence on the SFR. 
In Fig.~\ref{fig:Lx_SFR}, we present the measured X-ray luminosity, $L_{0.2-2.3keV}$, versus the SFR estimated using the broadband SED fitting method for our sample of  SFGs. 
As already found in previous works, we find a positive correlation between X-ray luminosity and SFR. 
 We perform a fit of our sample of SFG using the linear model:
\begin{equation}
   \rm{ log(L_x)=A + B \cdot log(SFR)},
    \label{Eq:Lx_SFR}
\end{equation}
where $L_x$ is in units of erg $s^{-1}$ and SFR is in units of $M_{\odot}$ $yr^{-1}$. 
We derive the fitting constants $A= 40.67 \pm 0.21$ and $B=0.57 \pm 0.20$.
Despite the correlation between the two parameters, due to the scatter of the sources, the fit does not yield statistically robust results, with a $\chi_r^2=11.32$.
The scaling relation from \cite{Lehmer2016} is plotted as a dashed black line representing the XRBs emission of a sample of normal galaxies in the local universe ($z \sim 0$). 
This sample was obtained as a combination of local normal galaxies and stacked sub-samples of normal galaxies in the $\sim7$Ms Chandra Deep Field-South (CDF-S) survey (\citealt{Luo2017}). The local normal galaxies subset analyzed by \cite{Lehmer2016} was observed at rest-frame emissions above 2keV. 
Therefore, \cite{Lehmer2016} corrected for the 0.5-2 keV emission range and added the hot gas contribution, which was determined based on the findings of \cite{Mineo2012}. 
Moreover, the CDF-S stacked sub-samples were generated based on the observed frame 0.5-1 keV emission, which probes the rest-frame 0.5-2 keV band emission and includes the total hot gas and XRB emission.
We notice that the majority of our SFGs lie above the Lx/SFR relation found by \cite{Lehmer2016}, but the trend of the relation seems to be very similar. 
This is reflected in the fitting parameters, with $B$ being consistent with the slope found in \cite{Lehmer2016} at $0.27\sigma,$ while $A$ is found to be not consistent at $4.3\sigma$. 

To explain the observed scatter of the SFGs sample from the literature relation, we took into account two possible scenarios:~1)~the scatter is due to differences in the intrinsic properties of the sources, such as different metallicity, intrinsic X-ray absorption or contribution from LMXBs and hot gas that do not follow the empirical relations employed in Section~\ref{sec:substraction}. 2)~the scatter is a consequence of the eROSITA sensitivity limit, shallower compared to \cite{Lehmer2016} at fixed redshift, which could preclude the detection of low Lx/SFR sources at higher redshift, resulting in a different normalization of the scaling relation. We stress that this different normalization is unlikely to be related to the SFR estimates, as \cite{Lehmer2016} adopt an UV+IR SFR tracer, compatible with the one from SED fitting employed in this work. 
Indeed, in Fig.~\ref{fig:Lx_SFR} we can notice an alleged dependence of the Lx/SFR relation on redshift, as $L_{0.2-2.3keV}/SFR$ seems to increase going to higher redshift, at fixed SFRs. 
Such dependence was indeed already observed and predicted, as we expected an evolution of HMXBs and LMXBs populations with cosmic time (\citealt{Basu2013a}, \citealt{Lehmer2016}). 
However, the evolution found in these works is only significant at $z>1$ and is not steep enough to address the rapid increase in the Lx/SFR with redshift observed in this work.

To address the first possibility, we estimated the metallicity of the sources having counterparts in the MPA/JHU catalog, using \cite{Tremonti2004} calibration. Indeed, we find that the sources with higher $L_{0.2-2.3keV}$ have a lower metallicity (of about 0.6 dex), as expected due to HMXBs being more  numerous  and  more  luminous  with  decreasing  metallicity, since weaker stellar winds allow more mass retention and tighter binary orbits, as demonstrated in X-ray binary population synthesis models (\citealt{Linden2010}, \citealt{Fragos2013}, \citealt{Basu2016}).  Unfortunately, as only four sources have SDSS counterparts, this result is not statistically robust, motivating the need for a full spectroscopic follow-up of the galaxies observed by the eFEDS survey.

\begin{figure}[]
    \centering
    \includegraphics[width=\hsize]{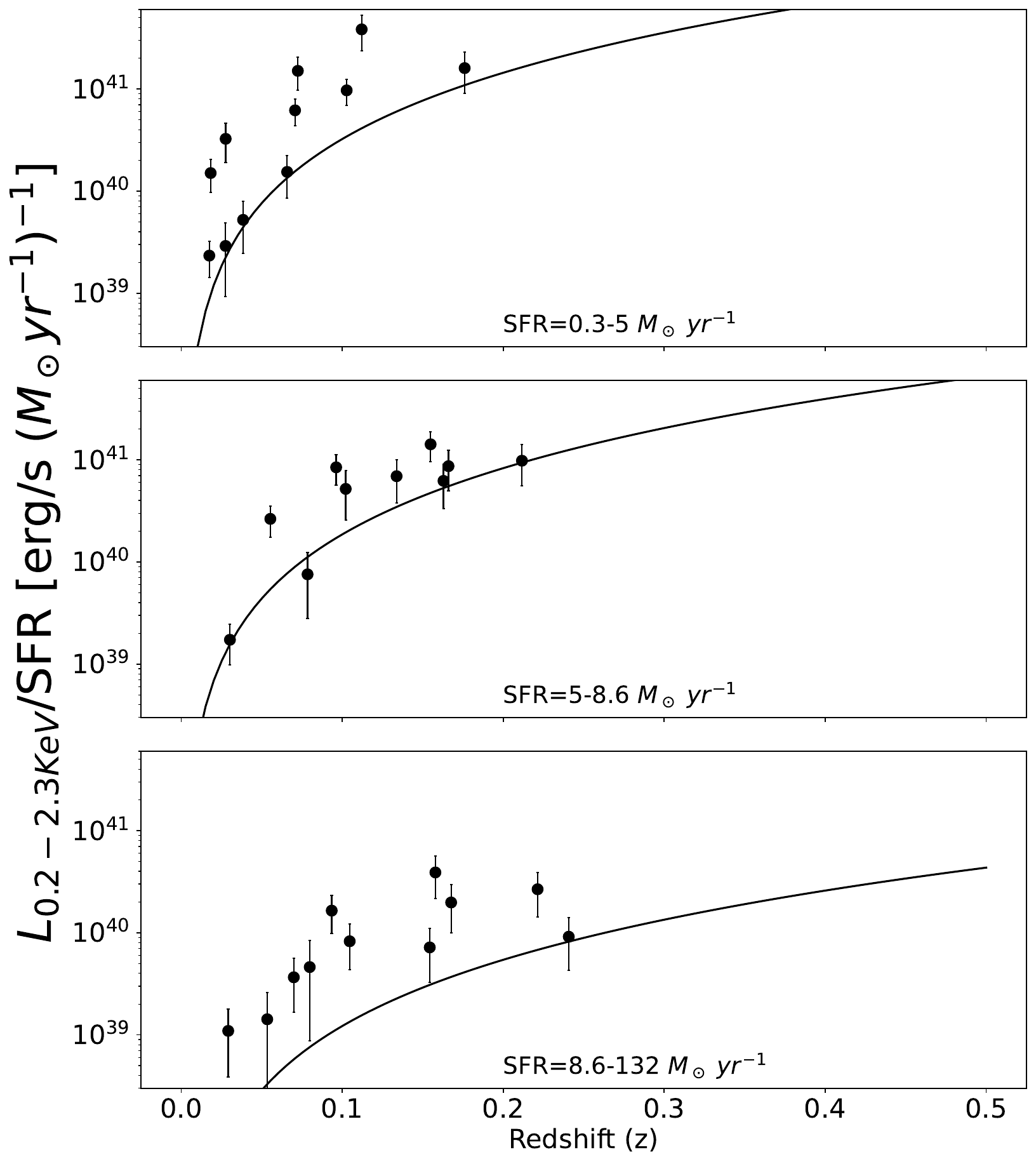}%
    \caption{ X-ray luminosity per  SFR unit ($L_{0.2-2.3keV}$/SFR) in the function of redshift for the sample of SFGs.  The solid black line represents the X-ray luminosity sensitivity limit of the eFEDS sample, rescaled by the max value of the SFR in each interval. }
    \label{fig:Lx_SFRz}
\end{figure}

Concerning the possibility that the observed difference might be due to the sensitivity limit of the eFEDS survey, we explored the evolution of the Lx/SFR scaling relation with redshift. In Fig.~\ref{fig:Lx_SFRz}, we show the $L_{0.2-2.3keV}$/SFR ratio as a function of redshift in three SFR intervals. 
We observe a considerable increase in the Lx/SFR ratio with increasing redshift, much steeper than the one found previously in the literature (\citealt{Lehmer2016}). However, we notice that the trend found in this work very well follows the X-ray luminosity limit of the eFEDS sample, represented in the figure by the solid black line. This confirms that our results are affected by completeness biases.

\begin{figure}[]
    \centering
    \includegraphics[width=\hsize]{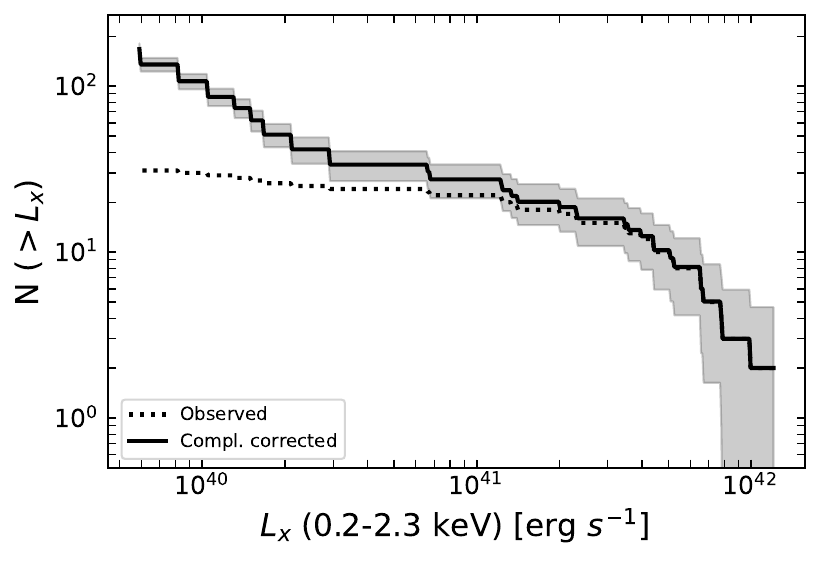}
    \includegraphics[width=\hsize]{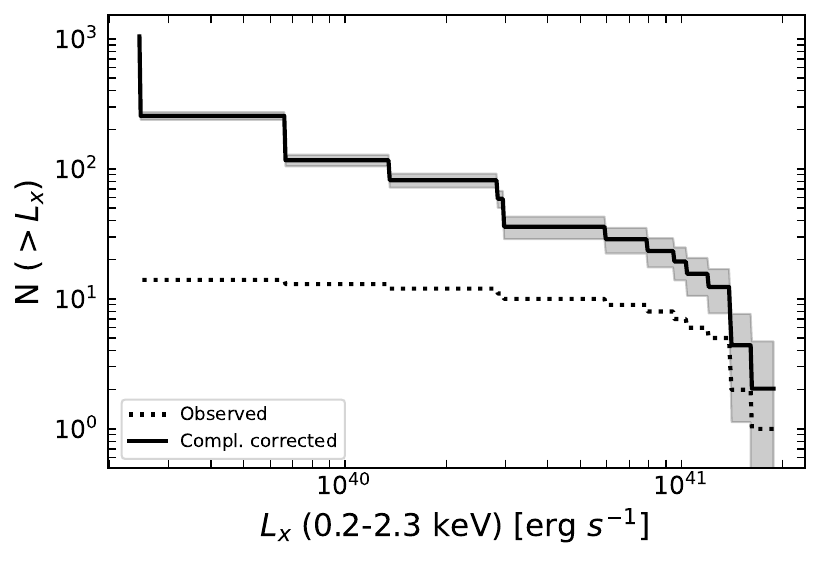}
    \caption{ Cumulative X-ray luminosity function for the sample of SF (top panel) and quiescent galaxies (bottom panel). The black solid line represents the completeness-corrected XLF, while the dotted line shows the observed
    XLF. The shaded region represents the 1$\sigma$ error.}
    \label{fig:completeness}
\end{figure}

\begin{figure}[]
    \centering
    \includegraphics[width=\hsize]{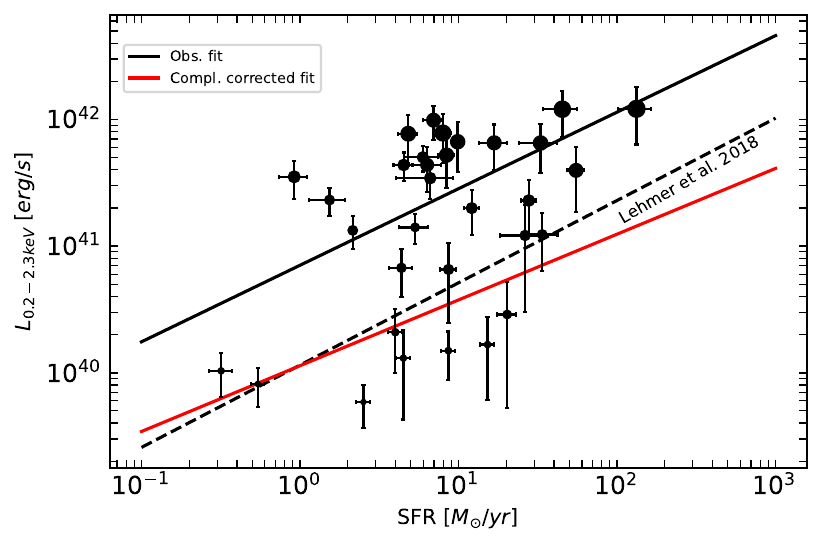}
    \caption{ X-ray luminosity in the 0.2-2.3 keV band as a function of the SFR. The size
    of the dots is proportional to the redshift of the sources. The red line shows the linear fit corrected for completeness, while the black line is the observed fit already shown in Fig.~\ref{fig:Lx_SFR}.}
    \label{fig:Lx_SFR_CORR}
\end{figure}

To address this problem, we correct the results for completeness. To do that, we make use of the MPA/JHU catalog to identify SFGs in our FoV that do not have X-ray counterparts in the eFEDS catalog. 
To maintain consistency with both with the methodology employed in \cite{Lehmer2016} and with the selection performed in this work, 
we select 2\,227 SFGs and 2\,367 passive galaxies from the MPA/JHU catalog using the selection criteria discussed in Section \ref{sec:MS}. In order to correct for completeness,  we weigh the X-ray sources according to the fraction of galaxies selected with the MPA/JHU catalog, in which they could have been detected. During this process, we removed the source having the lowest X-ray luminosity in our sample from the fitting
procedure, given it is the only source observed at that luminosity and on the verge of the eROSITA sensitivity limit. This source, which is attributed a weight of two orders of magnitude greater than the rest of the sources, does not represent a sufficient statistic to ensure a reliable correction for completeness, thus we do not consider it in the process. In Fig. \ref{fig:completeness} we show the observed and the completeness-corrected X-ray luminosity function (XLF) for our sample of SF and quiescent galaxies. We can notice that the observed XLF flattens due to completeness already at $L_{0.2-2.3 keV}\sim 10^{41}$ erg/s. Figure \ref{fig:Lx_SFR_CORR} shows the  $L_{0.2-2.3 keV}$-SFR scaling relation corrected for completeness (red solid line). We derived the fitting constants for the completeness-corrected curve as $A_{c.c}= 40.05 \pm 0.05$ and $B_{c.c}=0.52 \pm 0.06$, consistent with the one found by \cite{Lehmer2016} at 0.05$\sigma$ and 1.76$\sigma,$ respectively.

\section{Lx-sSFR relation}
In the previous section, we discussed the relation between the X-ray luminosity and star formation activity for the sample of 34 SFGs detected in the eFEDS FoV. However, as discussed in Section~\ref{sec:Introduction}, the X-ray emission of normal galaxies is not only dominated by the contribution of HMXBs, which is expected to scale with the SFR, but also by the contribution of LMXBs, which is expected to scale with the $M_{star}$. It was shown that the ratio of HMXB-to-LMXB emission is sensitive to the specific SFR (sSFR) and can be quantified with the scaling factors $\alpha \equiv L_{x,LMXB}/M_{star}$ and $\beta \equiv L_{x,HMXB}/SFR$, obtained as fitting constants of the empirical relation in the form (\citealt{Mineo2014}, \citealt{Lehmer2016}):
\begin{equation}
 \rm{   L_x = \alpha M_{star} + \beta SFR}.
    \label{Eq:Lx_sSFR}
\end{equation}

\begin{figure*}[]
    \centering
    \includegraphics[width=0.8\hsize]{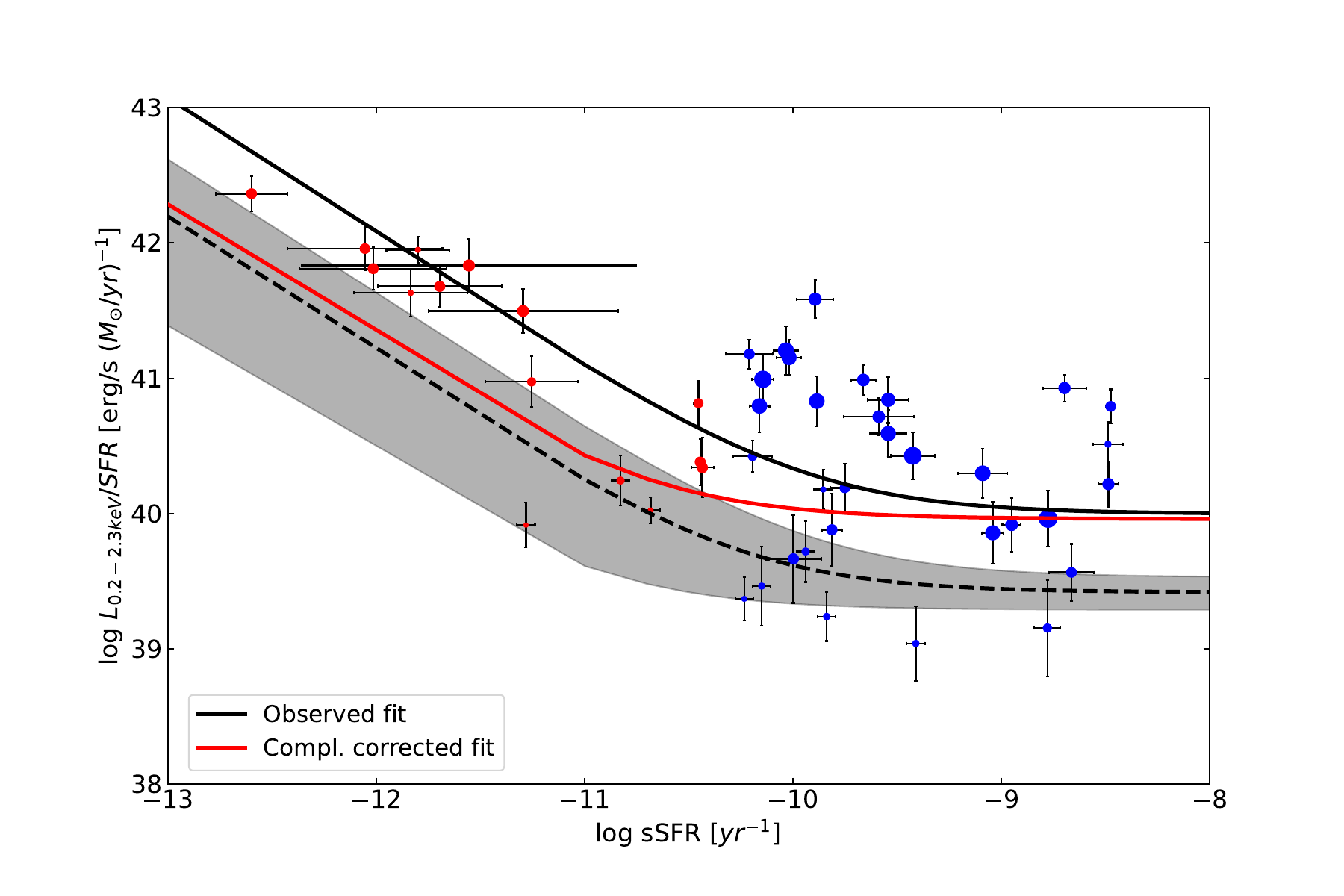}
    
    \caption{ X-ray luminosity in the 0.2-2.3 keV band scaled by the SFR in the function of the sSFR for the full sample of normal galaxies, selected in this work, both quiescent (red dots) and SFGs (blue dots). The solid black line represents the observed fit of the sources, while the red solid line represents the completeness corrected fit. \cite{Lehmer2016} fit at $z = 0.09$ is pictured as a black dashed line, with the shaded region representing the $3\sigma$ dispersion. The size
    of the dots scales with the redshift of the sources. 
    \ref{Sec:Results}.}
    \label{fig:Lx_sSFR}
\end{figure*}

Figure~\ref{fig:Lx_sSFR} shows the $L_{0.2-2.3keV}$ as a function of the sSFR for our sample of normal galaxies. 
We stress that for this plot we only removed the hot gas, AB, and CV components following the process described in Section~\ref{sec:substraction} and leaving the HMXBs and LMXBs emissions untouched. 
The scaling relation of \cite{Lehmer2016} at the mean redshift of our sample ($z \sim 0.09$) is plotted as the dashed black line, with dispersion in gray. 
Figure~\ref{fig:Lx_sSFR} shows a large dispersion from the relation, both for quiescent and SF galaxies. 
In the low-sSFR end, we do not notice any trend of the scatter with the redshift, having the sources from the entire redshift range clumped in the same region. 
Nevertheless, considering the large uncertainties on the Lx/SFR and sSFR, the sources in the low-sSFR regime are consistent with the locus of the scaling relation. 
For the SFGs, the scatter is much more accentuated and depends on redshift as already observed for the Lx-SFR scaling relation. 
In Fig.~\ref{fig:Lx_sSFR} we show both the observed and the completeness-corrected fits, as  black and red solid lines, respectively. 
We can notice that correcting for completeness lowers the relation, but in the high sSFR end, the fit still lies above the one previously observed by \cite{Lehmer2016}. 
For the completeness-corrected fit, we estimate the fitting parameters $\alpha=29.25 \pm 0.08$ and $\beta=39.95 \pm 0.03$.
Thus, correcting for completeness we obtain $\alpha$ consistent with the literature value, but we still observe a higher normalization for the SFGs, resulting in the $\beta$ parameter not consistent with what was previously found. 

\begin{figure}[]
    \centering
    \includegraphics[width=1\hsize]{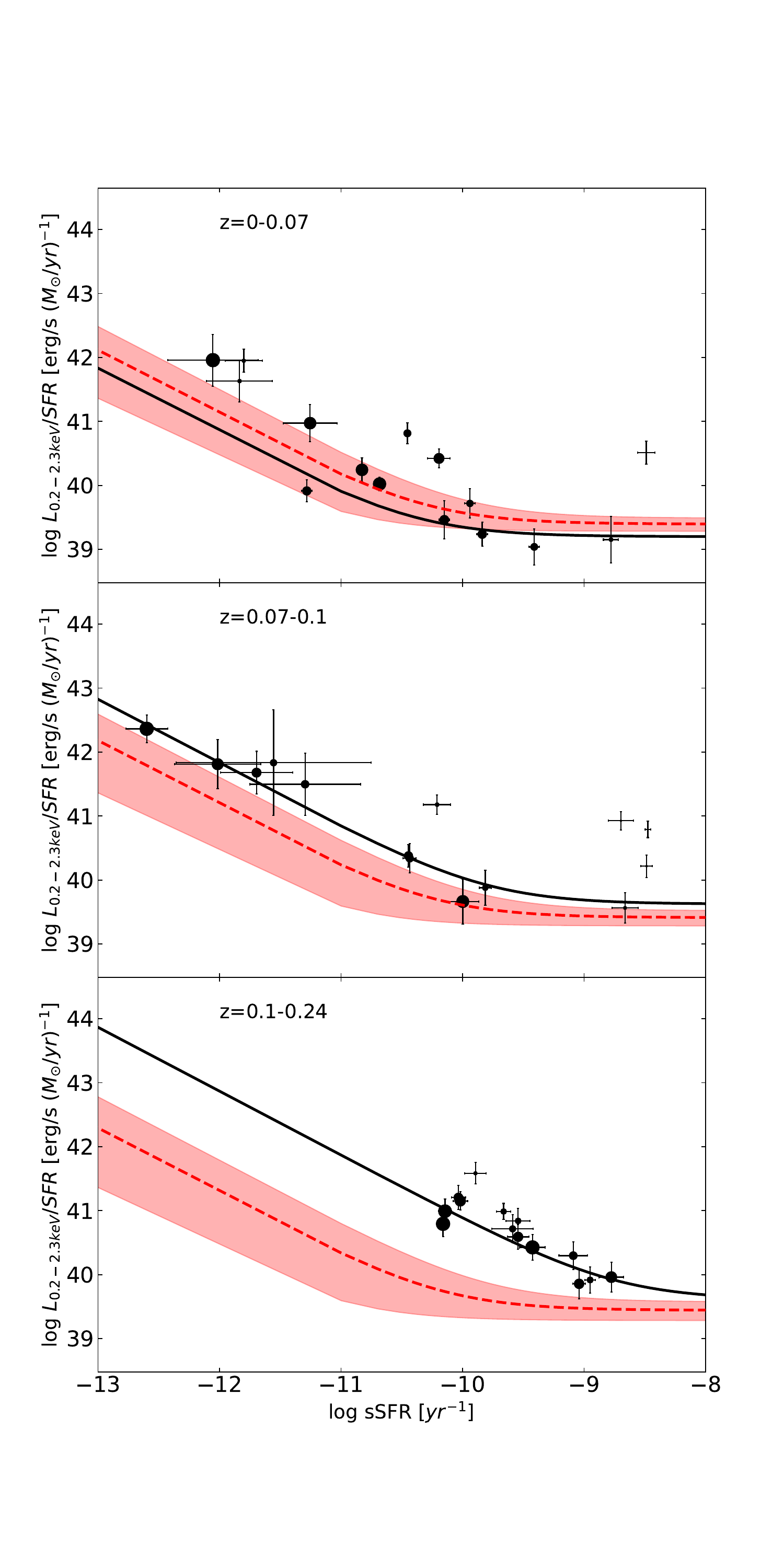}
        \caption{ X-ray luminosity in the 0.2-2.3 keV band scaled by the SFR in the function of the sSFR for the full sample of normal galaxies divided into three redshift bins. The points size is proportional to the $M_{star}$. The solid black lines represent the observed fit of the sources, while the red dashed lines represent \cite{Lehmer2016} relation estimated at the mean value of the redshift for each bin. The shaded red region represents the $3\sigma$ dispersion.}
    \label{fig:Lx_sSFR_z}
\end{figure}

To understand the role of the redshift on the estimated fitting parameters, in Fig. \ref{fig:Lx_sSFR_z}, we show the $L_{0.2-2.3keV}$/SFR as a function of the sSFR for three redshift bins. 
We divided our sample in order to have the same number of objects in each bin. 
We notice that for the lowest redshift bin, where the sample is the most complete, our fitted relation is consistent with what was found previously in the literature. Given it is incomplete in the highest ranges of redshift, we do not perform any statistical analysis on the evolution of the scaling relation with redshift. 
However, in the case where they are complete enough, we report the fitting parameters $\alpha=28.81 \pm 0.25$ and $\beta=39.19 \pm 0.3$ in the range of $z=0-0.07$. Table~\ref{table:fits} shows the best-fit parameters, corrected for completeness, for both the Lx-SFR and Lx/SFR-sSFR scaling relations.

\begin{table*}[]
\centering
\caption{Summary of the fits performed on the completeness corrected eFEDS sample.}
\label{table:fits}
\begin{tabular}{ccccc }
\hline
\textbf{Function} & \textbf{Parameter} & \textbf{Fitted value} & \textbf{z} & \textbf{Literature comparison (\citealt{Lehmer2016})} \\ \hline
 $log(L_x)=A + B \cdot log(SFR)$ 
 & A  & $40.05 \pm 0.05$ & 0 - 0.23 & $40.06 \pm 0.05$ \\
 & B  & $0.52 \pm 0.06$ & 0 - 0.23 & $0.65 \pm 0.04$\\ 
\\
$L_x = \alpha M_{star} + \beta SFR$ 
 & $\alpha$ & $29.25\pm 0.08$ & 0 - 0.23 & $29.04 \pm 0.17$\\
 &          & $28.81\pm 0.25$ & 0 - 0.07 \\
 \\
  & $\beta$  & $39.95 \pm 0.02$ & 0 - 0.23 & $39.66 \pm 0.03$\\
  &          & $39.19 \pm 0.03$ & 0 - 0.07\\
\\

\hline
\end{tabular}
\end{table*}

\begin{figure}[]
    \centering
    \includegraphics[width=1\hsize]{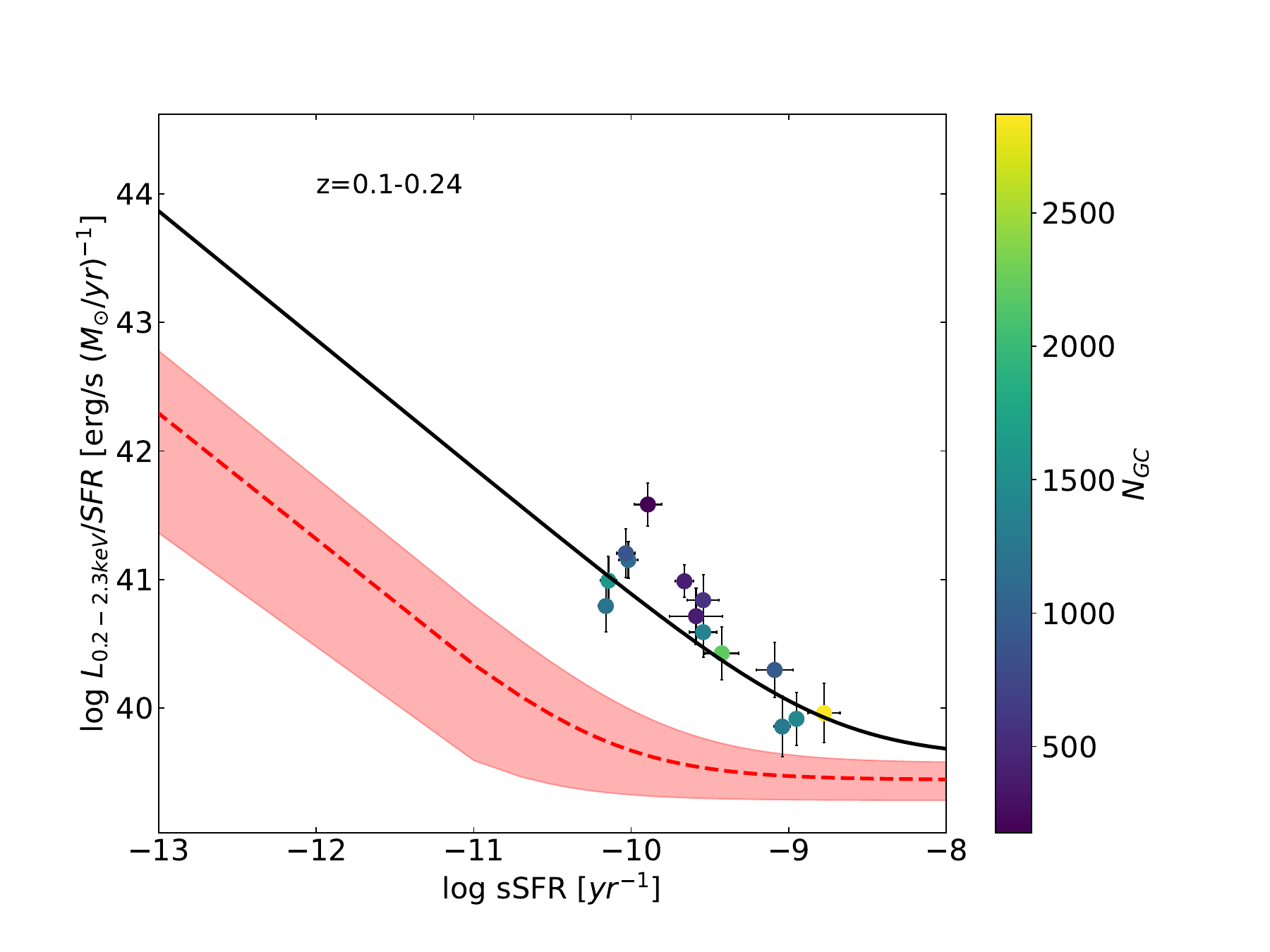}
    
    \caption{X-ray luminosity in the 0.2-2.3 keV band scaled by the SFR as a function of the sSFR, color-coded by the number of globular clusters. The lines are the same as Fig.~\ref{fig:Lx_sSFR_z}.}
    \label{fig:GC}
\end{figure}

We stress that for each panel, the plotted literature relation is estimated according to the mean value of redshift in the bin. Thus, even considering the completeness, it is interesting to notice that all objects in the highest redshift bin are clustered above the scaling relation.
\citealt{Vulic2022} found a similar trend for a sample of low redshift SFGs detected by the Heraklion Extragalactic Catalogue (HECATE) in the eFEDS field. 
They found that high sSFR dwarf galaxies tended to have higher values of the $L_{0.5-2.0keV}/SFR$ than expected by the scaling relation. 
For this reason, in Fig.~\ref{fig:Lx_sSFR_z}, we show the scaling of the size of the symbols according to their $M_{star}$.  In the redshift range z=0.07-0.1 we notice a slight preference of less massive galaxies ($\sim 10^{10}$ $M_{\odot}$) to scatter from the relation, at high sSFR. However, we do not find any statistically significant trend with the size of the galaxies, with all of our sources having  comparable $M_{star}$, distributed between $10^{10}$-$10^{11}$ $M_{\odot}$. 
Thus, the scatter of the sources at higher redshift should be traced back to other reasons, such as metallicity differences or an enhanced contribution of LMXBs, that could severely affect the X-ray luminosity of SFGs. To address the first possibility, as already discussed in the previous section, an accurate spectral analysis is necessary. Instead, regarding the second possibility, one explanation of this hypothetical enhanced contribution may be the presence of a large population of globular clusters (GC). In fact, it is known that the formation of LMXBs in GCs is favored as the high stellar density near the center of GCs may trigger the formation of binaries either by three-body process or by tidal capture. This component is usually not taken into account in theoretical XRB population-synthesis models. To have a rough idea of the GC population of our sample of galaxies, we use the empirical relation presented in \cite{Harris2013}, which relates the V-band absolute magnitude to the total number of GCs (see their Eq. 4). The relation is in the form:

\begin{equation}
    SN \equiv N_{GC} \times 10^{0.4(M_V^T+15)}
    \label{Eq:GC}
\end{equation}

where SN is the specific frequency of GCs and $M_V^T$ is the absolute magnitude in the V-band. They calibrated this relation on a sample of 422 sources, composed by elliptical, spiral, and irregular galaxies.
As we do not have measurements of the specific frequency of GCs, we assume $S/N=1$, as the V-band luminosity range of our sample lies in the region where the "U" shaped relation flattens to unity (see Fig. 10 in \citealt{Harris2013}). 
In Fig.~\ref{fig:GC}, we again show  the same subsample presented in the bottom panel of Fig.~\ref{fig:Lx_sSFR_z}, but color-coded by the number of GCs estimated with the  formula from \cite{Harris2013}. We do not notice any significant increase in the $L_{0.2-2.3keV}/SFR$ according to different GC populations. 
The same result is found for the other two redshift ranges. 

In conclusion, it is clear that a statistical analysis on normal galaxies carried out with eROSITA will be inevitably affected by severe completeness biases. Thus, to perform an accurate study of the evolution of the XRB contribution to the X-ray emission of galaxies is essential to perform a stacking process, as already performed in previous works (\citealt{Lehmer2016}) or any other method that addresses the X-ray non-detections. In this way, we would swap in information about the sources to have a more accurate statistical sample of faint X-ray sources, which would be able to populate the low $L_{0.2-2.3keV}/SFR$ at higher redshift. A straightforward follow-up of this work is geared in this direction. 

\section{Summary and conclusions}
\label{Sec:Conclusions}

We performed an analysis of the X-ray properties of a sample of normal galaxies with a negligible AGN contribution observed by SRG/eROSITA for the eROSITA Final Equatorial Depth Survey. 
The main goal of this work is to explore the contribution to the total X-ray emission given by HMXBs and LMXBs, and how it scales in terms of SFR and $M_{star}$. 
For this purpose, we made use of X-ray photometry in the 0.2-2.3 keV band for a sample of 27\,369 sources in the eFEDS field (\citealt{Brunner2022}). In order to estimate the physical parameters, we make use of ancillary data from the UV to the MIR provided in \cite{Salvato2022} to fit the SED of the galaxies, using the CIGALE code (Section~\ref{Sec:Methodology}).
To guarantee the quality of the fit and to remove possible foreground Galactic sources, we performed several quality cuts (discussed in Section~\ref{Sec:Data}) that allowed us to narrow the sample down to 888 galaxies. 

To ensure the reliability of the SFR estimates we validated our results (when possible) using other indicators. To do so, we cross-matched our sample with the \textit{Herschel Extragalactic Legacy Project} (HELP) survey in the GAMA09 field and with MPA/JHU catalog based on the Sloan Digital Sky Survey DR7 release (\citealt{Abazajian2009}). 
In this way, we acquired FIR and spectral data for 48 and 34 sources, respectively. We found consistent estimates with the SFR resulting from the fit of the SED up to SPIRE FIR photometry. On the other hand, we found a consistent difference between the SFR estimated with the $H_\alpha$ line and the one resulting from the fit of the SED up to MIR photometry. Making use of the BPT diagram to classify these objects, we found the difference to be driven by LINER and Seyfert galaxies, which are the source of the $H_\alpha$ emission for the low-SFR sources. At the end of the process, we updated the physical properties of 48 galaxies with the results from the fit of the SED up to the FIR, and the SFR of five galaxies selected by the BPT diagram as star-forming with the SFR from the $H_\alpha$ line.

To isolate the contribution of XRBs we substracted the X-ray emission from hot gas, CVs, and ABs. For quiescent galaxies, we used the K-band luminosity to estimate the contribution from hot gas, following the prescription discussed in \cite{Civano2014}. For SFGs, we employ the relation between $L_{x,LMXBs}$ and $M_*$ found by \cite{Gilfanov2004} to estimate the contribution from LMXBs (Eq.~\ref{Eq:LMXBs}). We account to the hot gas using \cite{Mineo2012} (Eq.~\ref{Eq:hot}). For both types, we use the relation from \cite{Boroson2011} to account for CVs and ABs (Eq~\ref{Eq:boroson}). These contributions are subtracted from the observed X-ray luminosity to accordingly isolate the emission from HMXBs and LMXBs.

After removing the contribution from hot gas, ABs, and CVs to study the properties of galaxies for which the X-ray
emission is dominated by XRBs we need to reveal the presence
of  non-stellar nuclear emission. To achieve this, we used a combination of observed photometry in the X-ray, optical, and MIR ranges, together with a selection based on the SED fitting. The criteria to select AGNs can be summarized as follows:

\begin{itemize}
    \item $L_{0.2-2.3keV} \geq 3\times10^{42}$ erg/s;
    \item X-ray-to-optical flux ratio of $\log(f_X/f_r)>-1$;
    \item X-ray-to-NIR flux ratio of $\log(f_X/f_{Ks})>-1.2$;
    \item MIR WISE photometry selection described in \cite{Assef2013};
    \item AGN contribution to the total X-ray emission estimated with the SED fitting $fAGN_{0.2-2.3keV}<0.1$.
\end{itemize}

At the end of the process, we were left with the final sample of 49 normal galaxies: 34 SFGs and 15 quiescent galaxies.

To study the HMXBs contribution to the X-ray luminosity we measured the constants A and B of the empirical relation 
already found in literature between $L_{0.2-2.3keV}$ and SFR (in the form of Eq.~\ref{Eq:Lx_SFR}). We stress that for this analysis, we subtract the LMXBs, hot gas, ABs, and CVs contributions to the total X-ray emission. We derived the fitting constants $A = 40.67 \pm 0.21$ and $B=0.57 \pm 0.20$. Despite the correlation, the fit did not yield statistically robust results. We found that the majority of our SFGs lie above the Lx/SFR relation found previously in the literature (\citealt{Lehmer2016}). This is reflected in the fitting parameter A, found to be inconsistent with previous measurements at $4.3\sigma$. To investigate this result, we discussed the possibility of a dependence on the completeness limit of the eFEDS survey, which could preclude the detection of
low Lx/SFR sources at higher redshift. Correcting for completeness using SFGs detected in the MPA/JHU catalog, we found very good agreement between the completeness corrected fitting constants and the literature. We derived $A_{c.c} = 40.05 \pm 0.05$ and $B_{c.c}=0.52 \pm 0.06$, consistent at $0.05\sigma$ and $1.76\sigma,$ respectively, with previous measurements. We conclude that the overall connection between X-ray luminosity and SFR of our population of SFGs is highly biased by completeness issues, but the scatter of the sources from the literature relation can be traced back to physical differences between the galaxies, such as metallicity differences, LMXBs contribution, or intrinsic X-ray absorption.

In order to study the ratio of HMXB-to-LMXB emission, which was shown to scale with the sSFR, we quantified the scaling factors $\alpha \equiv L_{x,LMXB}/M_{star}$ and $\beta \equiv L_{x,HMXB}/SFR$, fitting the empirical relation presented in Eq.~\ref{Eq:Lx_sSFR}. For this analysis, for both quiescent and SFGs, we subtract the hot gas, AB, and CV contributions to the X-ray luminosity. Correcting the full sample of normal galaxies for completeness, we derived the fitting parameters $\alpha=29.25 \pm 0.08$ and $\beta=39.95 \pm 0.02$. Thus, correcting for completeness is not enough to address the scatter from the relation at high sSFR, resulting in a $\beta$ parameter that is not consistent with the values found previously in the literature. Nonetheless, we found that the scatter of the sources from the relation is mainly carried out by high redshift sources, concluding that the statistical trend of the empirical relation is highly biased by completeness. Indeed, for the lowest redshift range where we are the most complete, we found consistent results with the literature. 
We address the scatter of the sources from the literature relation at high redshift by discussing the possibility of an enhanced LMXBs contribution due to an overpopulation of GCs, which would favor the formation of binary systems. We address this possibility by estimating the expected number of GCs employing the relation presented in \cite{Harris2013} (Eq.~\ref{Eq:GC}). We do not find any statistical correlation between the Lx/SFR and the number of GCs. We concluded that an accurate study
of the evolution of the XRB contribution to the X-ray emission of galaxies carried out with eROSITA must be performed, accounting for X-ray non-detections, to overcome the severe completeness biases. Furthermore, to investigate the scatter of the sources from the predicted relations, a full spectral follow-up of the eFEDS survey is necessary. 
Future works must be carried out with these views in mind.

 \begin{acknowledgements}
We would like to thank Michele Cantiello and Bret Lehmer for helping the realization of this project with insightful discussions. 
GR, KM, MH, and Junais acknowledge support from the National Science Centre (UMO-2018/30/E/ST9/00082). FP gratefully acknowledges support from the Polish National Science Centre grant (UMO-2018/30/M/ST9/00757).
MB gratefully acknowledges support from the ANID BASAL project FB210003 and from the FONDECYT regular grant 1211000.

 \end{acknowledgements}

\bibliographystyle{aa} 
\bibliography{bibliography}

\begin{thebibliography}{85}
\expandafter\ifx\csname natexlab\endcsname\relax\def\natexlab#1{#1}\fi

\bibitem[{{Abazajian} {et~al.}(2009){Abazajian}, {Adelman-McCarthy},
  {Ag{\"u}eros}, {Allam}, {Allende Prieto}, {An}, {Anderson}, {Anderson},
  {Annis}, {Bahcall}, {Bailer-Jones}, {Barentine}, {Bassett}, {Becker},
  {Beers}, {Bell}, {Belokurov}, {Berlind}, {Berman}, {Bernardi}, {Bickerton},
  {Bizyaev}, {Blakeslee}, {Blanton}, {Bochanski}, {Boroski}, {Brewington},
  {Brinchmann}, {Brinkmann}, {Brunner}, {Budav{\'a}ri}, {Carey}, {Carliles},
  {Carr}, {Castander}, {Cinabro}, {Connolly}, {Csabai}, {Cunha}, {Czarapata},
  {Davenport}, {de Haas}, {Dilday}, {Doi}, {Eisenstein}, {Evans}, {Evans},
  {Fan}, {Friedman}, {Frieman}, {Fukugita}, {G{\"a}nsicke}, {Gates},
  {Gillespie}, {Gilmore}, {Gonzalez}, {Gonzalez}, {Grebel}, {Gunn},
  {Gy{\"o}ry}, {Hall}, {Harding}, {Harris}, {Harvanek}, {Hawley}, {Hayes},
  {Heckman}, {Hendry}, {Hennessy}, {Hindsley}, {Hoblitt}, {Hogan}, {Hogg},
  {Holtzman}, {Hyde}, {Ichikawa}, {Ichikawa}, {Im}, {Ivezi{\'c}}, {Jester},
  {Jiang}, {Johnson}, {Jorgensen}, {Juri{\'c}}, {Kent}, {Kessler}, {Kleinman},
  {Knapp}, {Konishi}, {Kron}, {Krzesinski}, {Kuropatkin}, {Lampeitl},
  {Lebedeva}, {Lee}, {Lee}, {French Leger}, {L{\'e}pine}, {Li}, {Lima}, {Lin},
  {Long}, {Loomis}, {Loveday}, {Lupton}, {Magnier}, {Malanushenko},
  {Malanushenko}, {Mandelbaum}, {Margon}, {Marriner}, {Mart{\'\i}nez-Delgado},
  {Matsubara}, {McGehee}, {McKay}, {Meiksin}, {Morrison}, {Mullally}, {Munn},
  {Murphy}, {Nash}, {Nebot}, {Neilsen}, {Newberg}, {Newman}, {Nichol},
  {Nicinski}, {Nieto-Santisteban}, {Nitta}, {Okamura}, {Oravetz}, {Ostriker},
  {Owen}, {Padmanabhan}, {Pan}, {Park}, {Pauls}, {Peoples}, {Percival}, {Pier},
  {Pope}, {Pourbaix}, {Price}, {Purger}, {Quinn}, {Raddick}, {Re Fiorentin},
  {Richards}, {Richmond}, {Riess}, {Rix}, {Rockosi}, {Sako}, {Schlegel},
  {Schneider}, {Scholz}, {Schreiber}, {Schwope}, {Seljak}, {Sesar}, {Sheldon},
  {Shimasaku}, {Sibley}, {Simmons}, {Sivarani}, {Allyn Smith}, {Smith},
  {Smol{\v{c}}i{\'c}}, {Snedden}, {Stebbins}, {Steinmetz}, {Stoughton},
  {Strauss}, {SubbaRao}, {Suto}, {Szalay}, {Szapudi}, {Szkody}, {Tanaka},
  {Tegmark}, {Teodoro}, {Thakar}, {Tremonti}, {Tucker}, {Uomoto}, {Vanden
  Berk}, {Vandenberg}, {Vidrih}, {Vogeley}, {Voges}, {Vogt}, {Wadadekar},
  {Watters}, {Weinberg}, {West}, {White}, {Wilhite}, {Wonders}, {Yanny},
  {Yocum}, {York}, {Zehavi}, {Zibetti}, \& {Zucker}}]{Abazajian2009}
{Abazajian}, K.~N., {Adelman-McCarthy}, J.~K., {Ag{\"u}eros}, M.~A., {et~al.}
  2009, \apjs, 182, 543

\bibitem[{{Aihara} {et~al.}(2018){Aihara}, {Arimoto}, {Armstrong}, {Arnouts},
  {Bahcall}, {Bickerton}, {Bosch}, {Bundy}, {Capak}, {Chan}, {Chiba}, {Coupon},
  {Egami}, {Enoki}, {Finet}, {Fujimori}, {Fujimoto}, {Furusawa}, {Furusawa},
  {Goto}, {Goulding}, {Greco}, {Greene}, {Gunn}, {Hamana}, {Harikane},
  {Hashimoto}, {Hattori}, {Hayashi}, {Hayashi}, {He{\l}miniak}, {Higuchi},
  {Hikage}, {Ho}, {Hsieh}, {Huang}, {Huang}, {Ikeda}, {Imanishi}, {Inoue},
  {Iwasawa}, {Iwata}, {Jaelani}, {Jian}, {Kamata}, {Karoji}, {Kashikawa},
  {Katayama}, {Kawanomoto}, {Kayo}, {Koda}, {Koike}, {Kojima}, {Komiyama},
  {Konno}, {Koshida}, {Koyama}, {Kusakabe}, {Leauthaud}, {Lee}, {Lin}, {Lin},
  {Lupton}, {Mand elbaum}, {Matsuoka}, {Medezinski}, {Mineo}, {Miyama},
  {Miyatake}, {Miyazaki}, {Momose}, {More}, {More}, {Moritani}, {Moriya},
  {Morokuma}, {Mukae}, {Murata}, {Murayama}, {Nagao}, {Nakata}, {Niida},
  {Niikura}, {Nishizawa}, {Obuchi}, {Oguri}, {Oishi}, {Okabe}, {Okamoto},
  {Okura}, {Ono}, {Onodera}, {Onoue}, {Osato}, {Ouchi}, {Price}, {Pyo}, {Sako},
  {Sawicki}, {Shibuya}, {Shimasaku}, {Shimono}, {Shirasaki}, {Silverman},
  {Simet}, {Speagle}, {Spergel}, {Strauss}, {Sugahara}, {Sugiyama}, {Suto},
  {Suyu}, {Suzuki}, {Tait}, {Takada}, {Takata}, {Tamura}, {Tanaka}, {Tanaka},
  {Tanaka}, {Tanaka}, {Terai}, {Terashima}, {Toba}, {Tominaga}, {Toshikawa},
  {Turner}, {Uchida}, {Uchiyama}, {Umetsu}, {Uraguchi}, {Urata}, {Usuda},
  {Utsumi}, {Wang}, {Wang}, {Wong}, {Yabe}, {Yamada}, {Yamanoi}, {Yasuda},
  {Yeh}, {Yonehara}, \& {Yuma}}]{Aihara2018}
{Aihara}, H., {Arimoto}, N., {Armstrong}, R., {et~al.} 2018, \pasj, 70, S4

\bibitem[{{Aird} {et~al.}(2017){Aird}, {Coil}, \& {Georgakakis}}]{Aird2017}
{Aird}, J., {Coil}, A.~L., \& {Georgakakis}, A. 2017, \mnras, 465, 3390

\bibitem[{Aird {et~al.}(2019)Aird, Coil, \& Georgakakis}]{Aird2019}
Aird, J., Coil, A.~L., \& Georgakakis, A. 2019, Monthly Notices of the Royal
  Astronomical Society, 484, 4360

\bibitem[{{Arnouts} {et~al.}(1999){Arnouts}, {Cristiani}, {Moscardini},
  {Matarrese}, {Lucchin}, {Fontana}, \& {Giallongo}}]{Arnouts1999}
{Arnouts}, S., {Cristiani}, S., {Moscardini}, L., {et~al.} 1999, \mnras, 310,
  540

\bibitem[{{Assef} {et~al.}(2013){Assef}, {Stern}, {Kochanek}, {Blain},
  {Brodwin}, {Brown}, {Donoso}, {Eisenhardt}, {Jannuzi}, {Jarrett}, {Stanford},
  {Tsai}, {Wu}, \& {Yan}}]{Assef2013}
{Assef}, R.~J., {Stern}, D., {Kochanek}, C.~S., {et~al.} 2013, \apj, 772, 26

\bibitem[{{Baldwin} {et~al.}(1981){Baldwin}, {Phillips}, \&
  {Terlevich}}]{Baldwin1981}
{Baldwin}, J.~A., {Phillips}, M.~M., \& {Terlevich}, R. 1981, \pasp, 93, 5

\bibitem[{{Basu-Zych} {et~al.}(2016){Basu-Zych}, {Lehmer}, {Fragos},
  {Hornschemeier}, {Yukita}, {Zezas}, \& {Ptak}}]{Basu2016}
{Basu-Zych}, A.~R., {Lehmer}, B., {Fragos}, T., {et~al.} 2016, \apj, 818, 140

\bibitem[{{Basu-Zych} {et~al.}(2013){Basu-Zych}, {Lehmer}, {Hornschemeier},
  {Bouwens}, {Fragos}, {Oesch}, {Belczynski}, {Brandt}, {Kalogera}, {Luo},
  {Miller}, {Mullaney}, {Tzanavaris}, {Xue}, \& {Zezas}}]{Basu2013a}
{Basu-Zych}, A.~R., {Lehmer}, B.~D., {Hornschemeier}, A.~E., {et~al.} 2013,
  \apj, 762, 45

\bibitem[{{Bendo} {et~al.}(2002){Bendo}, {Joseph}, {Wells}, {Gallais}, {Haas},
  {Heras}, {Klaas}, {Laureijs}, {Leech}, {Lemke}, {Metcalfe}, {Rowan-Robinson},
  {Schulz}, \& {Telesco}}]{Bendo}
{Bendo}, G.~J., {Joseph}, R.~D., {Wells}, M., {et~al.} 2002, \aj, 124, 1380

\bibitem[{{Bianchi}(2014)}]{Bianchi2014}
{Bianchi}, L. 2014, \apss, 354, 103

\bibitem[{{Boquien} {et~al.}(2019){Boquien}, {Burgarella}, {Roehlly}, {Buat},
  {Ciesla}, {Corre}, {Inoue}, \& {Salas}}]{Boquien2019}
{Boquien}, M., {Burgarella}, D., {Roehlly}, Y., {et~al.} 2019, \aap, 622, A103

\bibitem[{{Boroson} {et~al.}(2011){Boroson}, {Kim}, \&
  {Fabbiano}}]{Boroson2011}
{Boroson}, B., {Kim}, D.-W., \& {Fabbiano}, G. 2011, \apj, 729, 12

\bibitem[{{Brunner} {et~al.}(2022){Brunner}, {Liu}, {Lamer}, {Georgakakis},
  {Merloni}, {Brusa}, {Bulbul}, {Dennerl}, {Friedrich}, {Liu}, {Maitra},
  {Nandra}, {Ramos-Ceja}, {Sanders}, {Stewart}, {Boller}, {Buchner}, {Clerc},
  {Comparat}, {Dwelly}, {Eckert}, {Finoguenov}, {Freyberg}, {Ghirardini},
  {Gueguen}, {Haberl}, {Kreykenbohm}, {Krumpe}, {Osterhage}, {Pacaud},
  {Predehl}, {Reiprich}, {Robrade}, {Salvato}, {Santangelo}, {Schrabback},
  {Schwope}, \& {Wilms}}]{Brunner2022}
{Brunner}, H., {Liu}, T., {Lamer}, G., {et~al.} 2022, \aap, 661, A1

\bibitem[{{Bruzual} \& {Charlot}(2003)}]{Bruzual2003}
{Bruzual}, G. \& {Charlot}, S. 2003, \mnras, 344, 1000

\bibitem[{{Buat} {et~al.}(2019){Buat}, {Ciesla}, {Boquien}, {Ma{\l}ek}, \&
  {Burgarella}}]{Buat2019}
{Buat}, V., {Ciesla}, L., {Boquien}, M., {Ma{\l}ek}, K., \& {Burgarella}, D.
  2019, \aap, 632, A79

\bibitem[{{Calzetti} {et~al.}(2000){Calzetti}, {Armus}, {Bohlin}, {Kinney},
  {Koornneef}, \& {Storchi-Bergmann}}]{Calzetti2000}
{Calzetti}, D., {Armus}, L., {Bohlin}, R.~C., {et~al.} 2000, \apj, 533, 682

\bibitem[{{Chabrier}(2003)}]{Chabrier2003}
{Chabrier}, G. 2003, \pasp, 115, 763

\bibitem[{{Ciesla} {et~al.}(2015){Ciesla}, {Charmandaris}, {Georgakakis},
  {Bernhard}, {Mitchell}, {Buat}, {Elbaz}, {LeFloc'h}, {Lacey}, {Magdis}, \&
  {Xilouris}}]{Ciesla2015}
{Ciesla}, L., {Charmandaris}, V., {Georgakakis}, A., {et~al.} 2015, \aap, 576,
  A10

\bibitem[{{Civano} {et~al.}(2014){Civano}, {Fabbiano}, {Pellegrini}, {Kim},
  {Paggi}, {Feder}, \& {Elvis}}]{Civano2014}
{Civano}, F., {Fabbiano}, G., {Pellegrini}, S., {et~al.} 2014, \apj, 790, 16

\bibitem[{{Dale} \& {Helou}(2002)}]{Dale2002}
{Dale}, D.~A. \& {Helou}, G. 2002, \apj, 576, 159

\bibitem[{{Dale} {et~al.}(2014){Dale}, {Helou}, {Magdis}, {Armus},
  {D{\'\i}az-Santos}, \& {Shi}}]{Dale2014}
{Dale}, D.~A., {Helou}, G., {Magdis}, G.~E., {et~al.} 2014, \apj, 784, 83

\bibitem[{{Dey} {et~al.}(2019){Dey}, {Schlegel}, {Lang}, {Blum}, {Burleigh},
  {Fan}, {Findlay}, {Finkbeiner}, {Herrera}, {Juneau}, {Landriau}, {Levi},
  {McGreer}, {Meisner}, {Myers}, {Moustakas}, {Nugent}, {Patej}, {Schlafly},
  {Walker}, {Valdes}, {Weaver}, {Y{\`e}che}, {Zou}, {Zhou}, {Abareshi},
  {Abbott}, {Abolfathi}, {Aguilera}, {Alam}, {Allen}, {Alvarez}, {Annis},
  {Ansarinejad}, {Aubert}, {Beechert}, {Bell}, {BenZvi}, {Beutler}, {Bielby},
  {Bolton}, {Brice{\~n}o}, {Buckley-Geer}, {Butler}, {Calamida}, {Carlberg},
  {Carter}, {Casas}, {Castander}, {Choi}, {Comparat}, {Cukanovaite}, {Delubac},
  {DeVries}, {Dey}, {Dhungana}, {Dickinson}, {Ding}, {Donaldson}, {Duan},
  {Duckworth}, {Eftekharzadeh}, {Eisenstein}, {Etourneau}, {Fagrelius},
  {Farihi}, {Fitzpatrick}, {Font-Ribera}, {Fulmer}, {G{\"a}nsicke},
  {Gaztanaga}, {George}, {Gerdes}, {Gontcho}, {Gorgoni}, {Green}, {Guy},
  {Harmer}, {Hernandez}, {Honscheid}, {Huang}, {James}, {Jannuzi}, {Jiang},
  {Joyce}, {Karcher}, {Karkar}, {Kehoe}, {Kneib}, {Kueter-Young}, {Lan},
  {Lauer}, {Le Guillou}, {Le Van Suu}, {Lee}, {Lesser}, {Perreault Levasseur},
  {Li}, {Mann}, {Marshall}, {Mart{\'\i}nez-V{\'a}zquez}, {Martini}, {du Mas des
  Bourboux}, {McManus}, {Meier}, {M{\'e}nard}, {Metcalfe},
  {Mu{\~n}oz-Guti{\'e}rrez}, {Najita}, {Napier}, {Narayan}, {Newman}, {Nie},
  {Nord}, {Norman}, {Olsen}, {Paat}, {Palanque-Delabrouille}, {Peng},
  {Poppett}, {Poremba}, {Prakash}, {Rabinowitz}, {Raichoor}, {Rezaie},
  {Robertson}, {Roe}, {Ross}, {Ross}, {Rudnick}, {Safonova}, {Saha},
  {S{\'a}nchez}, {Savary}, {Schweiker}, {Scott}, {Seo}, {Shan}, {Silva},
  {Slepian}, {Soto}, {Sprayberry}, {Staten}, {Stillman}, {Stupak}, {Summers},
  {Sien Tie}, {Tirado}, {Vargas-Maga{\~n}a}, {Vivas}, {Wechsler}, {Williams},
  {Yang}, {Yang}, {Yapici}, {Zaritsky}, {Zenteno}, {Zhang}, {Zhang}, {Zhou}, \&
  {Zhou}}]{Dey2019}
{Dey}, A., {Schlegel}, D.~J., {Lang}, D., {et~al.} 2019, \aj, 157, 168

\bibitem[{{Edge} {et~al.}(2013){Edge}, {Sutherland}, {Kuijken}, {Driver},
  {McMahon}, {Eales}, \& {Emerson}}]{Edge2013}
{Edge}, A., {Sutherland}, W., {Kuijken}, K., {et~al.} 2013, The Messenger, 154,
  32

\bibitem[{{Fabbiano}(2006)}]{Fabbiano2006}
{Fabbiano}, G. 2006, \araa, 44, 323

\bibitem[{{Fragos} {et~al.}(2013){Fragos}, {Lehmer}, {Naoz}, {Zezas}, \&
  {Basu-Zych}}]{Fragos2013}
{Fragos}, T., {Lehmer}, B.~D., {Naoz}, S., {Zezas}, A., \& {Basu-Zych}, A.
  2013, \apjl, 776, L31

\bibitem[{{Gilfanov}(2004)}]{Gilfanov2004}
{Gilfanov}, M. 2004, \mnras, 349, 146

\bibitem[{{Giovannoli} {et~al.}(2011){Giovannoli}, {Buat}, {Noll},
  {Burgarella}, \& {Magnelli}}]{Giovannoli2011}
{Giovannoli}, E., {Buat}, V., {Noll}, S., {Burgarella}, D., \& {Magnelli}, B.
  2011, \aap, 525, A150

\bibitem[{{Griffin} {et~al.}(2010){Griffin}, {Abergel}, {Abreu}, {Ade},
  {Andr{\'e}}, {Augueres}, {Babbedge}, {Bae}, {Baillie}, {Baluteau}, {Barlow},
  {Bendo}, {Benielli}, {Bock}, {Bonhomme}, {Brisbin}, {Brockley-Blatt},
  {Caldwell}, {Cara}, {Castro-Rodriguez}, {Cerulli}, {Chanial}, {Chen},
  {Clark}, {Clements}, {Clerc}, {Coker}, {Communal}, {Conversi}, {Cox},
  {Crumb}, {Cunningham}, {Daly}, {Davis}, {de Antoni}, {Delderfield}, {Devin},
  {di Giorgio}, {Didschuns}, {Dohlen}, {Donati}, {Dowell}, {Dowell}, {Duband},
  {Dumaye}, {Emery}, {Ferlet}, {Ferrand}, {Fontignie}, {Fox}, {Franceschini},
  {Frerking}, {Fulton}, {Garcia}, {Gastaud}, {Gear}, {Glenn}, {Goizel},
  {Griffin}, {Grundy}, {Guest}, {Guillemet}, {Hargrave}, {Harwit}, {Hastings},
  {Hatziminaoglou}, {Herman}, {Hinde}, {Hristov}, {Huang}, {Imhof}, {Isaak},
  {Israelsson}, {Ivison}, {Jennings}, {Kiernan}, {King}, {Lange}, {Latter},
  {Laurent}, {Laurent}, {Leeks}, {Lellouch}, {Levenson}, {Li}, {Li},
  {Lilienthal}, {Lim}, {Liu}, {Lu}, {Madden}, {Mainetti}, {Marliani}, {McKay},
  {Mercier}, {Molinari}, {Morris}, {Moseley}, {Mulder}, {Mur}, {Naylor},
  {Nguyen}, {O'Halloran}, {Oliver}, {Olofsson}, {Olofsson}, {Orfei}, {Page},
  {Pain}, {Panuzzo}, {Papageorgiou}, {Parks}, {Parr-Burman}, {Pearce},
  {Pearson}, {P{\'e}rez-Fournon}, {Pinsard}, {Pisano}, {Podosek}, {Pohlen},
  {Polehampton}, {Pouliquen}, {Rigopoulou}, {Rizzo}, {Roseboom}, {Roussel},
  {Rowan-Robinson}, {Rownd}, {Saraceno}, {Sauvage}, {Savage}, {Savini},
  {Sawyer}, {Scharmberg}, {Schmitt}, {Schneider}, {Schulz}, {Schwartz},
  {Shafer}, {Shupe}, {Sibthorpe}, {Sidher}, {Smith}, {Smith}, {Smith},
  {Spencer}, {Stobie}, {Sudiwala}, {Sukhatme}, {Surace}, {Stevens}, {Swinyard},
  {Trichas}, {Tourette}, {Triou}, {Tseng}, {Tucker}, {Turner}, {Vaccari},
  {Valtchanov}, {Vigroux}, {Virique}, {Voellmer}, {Walker}, {Ward}, {Waskett},
  {Weilert}, {Wesson}, {White}, {Whitehouse}, {Wilson}, {Winter}, {Woodcraft},
  {Wright}, {Xu}, {Zavagno}, {Zemcov}, {Zhang}, \& {Zonca}}]{Griffin2010}
{Griffin}, M.~J., {Abergel}, A., {Abreu}, A., {et~al.} 2010, \aap, 518, L3

\bibitem[{{Grimes} {et~al.}(2005){Grimes}, {Heckman}, {Strickland}, \&
  {Ptak}}]{Grimes2005}
{Grimes}, J.~P., {Heckman}, T., {Strickland}, D., \& {Ptak}, A. 2005, \apj,
  628, 187

\bibitem[{{Grimm} {et~al.}(2003){Grimm}, {Gilfanov}, \& {Sunyaev}}]{Grimm2003}
{Grimm}, H.~J., {Gilfanov}, M., \& {Sunyaev}, R. 2003, \mnras, 339, 793

\bibitem[{{Harris} {et~al.}(2013){Harris}, {Harris}, \& {Alessi}}]{Harris2013}
{Harris}, W.~E., {Harris}, G. L.~H., \& {Alessi}, M. 2013, \apj, 772, 82

\bibitem[{{Heckman}(1980)}]{Heckman}
{Heckman}, T.~M. 1980, \aap, 87, 152

\bibitem[{{Ilbert} {et~al.}(2006){Ilbert}, {Arnouts}, {McCracken},
  {Bolzonella}, {Bertin}, {Le F{\`e}vre}, {Mellier}, {Zamorani}, {Pell{\`o}},
  {Iovino}, {Tresse}, {Le Brun}, {Bottini}, {Garilli}, {Maccagni}, {Picat},
  {Scaramella}, {Scodeggio}, {Vettolani}, {Zanichelli}, {Adami}, {Bardelli},
  {Cappi}, {Charlot}, {Ciliegi}, {Contini}, {Cucciati}, {Foucaud}, {Franzetti},
  {Gavignaud}, {Guzzo}, {Marano}, {Marinoni}, {Mazure}, {Meneux}, {Merighi},
  {Paltani}, {Pollo}, {Pozzetti}, {Radovich}, {Zucca}, {Bondi}, {Bongiorno},
  {Busarello}, {de La Torre}, {Gregorini}, {Lamareille}, {Mathez}, {Merluzzi},
  {Ripepi}, {Rizzo}, \& {Vergani}}]{Ilbert2006}
{Ilbert}, O., {Arnouts}, S., {McCracken}, H.~J., {et~al.} 2006, \aap, 457, 841

\bibitem[{{Just} {et~al.}(2007){Just}, {Brandt}, {Shemmer}, {Steffen},
  {Schneider}, {Chartas}, \& {Garmire}}]{Just2007}
{Just}, D.~W., {Brandt}, W.~N., {Shemmer}, O., {et~al.} 2007, \apj, 665, 1004

\bibitem[{{Kennicutt}(1998)}]{Kennicutt98}
{Kennicutt}, Robert~C., J. 1998, \araa, 36, 189

\bibitem[{{Kim} \& {Fabbiano}(2013)}]{Kim2013}
{Kim}, D.-W. \& {Fabbiano}, G. 2013, \apj, 776, 116

\bibitem[{{Komatsu} {et~al.}(2011){Komatsu}, {Smith}, {Dunkley}, {Bennett},
  {Gold}, {Hinshaw}, {Jarosik}, {Larson}, {Nolta}, {Page}, {Spergel},
  {Halpern}, {Hill}, {Kogut}, {Limon}, {Meyer}, {Odegard}, {Tucker}, {Weiland},
  {Wollack}, \& {Wright}}]{Komatsu2011}
{Komatsu}, E., {Smith}, K.~M., {Dunkley}, J., {et~al.} 2011, \apjs, 192, 18

\bibitem[{{Kuijken} {et~al.}(2019){Kuijken}, {Heymans}, {Dvornik},
  {Hildebrandt}, {de Jong}, {Wright}, {Erben}, {Bilicki}, {Giblin}, {Shan},
  {Getman}, {Grado}, {Hoekstra}, {Miller}, {Napolitano}, {Paolilo}, {Radovich},
  {Schneider}, {Sutherland}, {Tewes}, {Tortora}, {Valentijn}, \& {Verdoes
  Kleijn}}]{Kuijken2019}
{Kuijken}, K., {Heymans}, C., {Dvornik}, A., {et~al.} 2019, \aap, 625, A2

\bibitem[{{Larkin} {et~al.}(1998){Larkin}, {Armus}, {Knop}, {Soifer}, \&
  {Matthews}}]{Larkin}
{Larkin}, J.~E., {Armus}, L., {Knop}, R.~A., {Soifer}, B.~T., \& {Matthews}, K.
  1998, \apjs, 114, 59

\bibitem[{{Lehmer} {et~al.}(2010){Lehmer}, {Alexander}, {Bauer}, {Brandt},
  {Goulding}, {Jenkins}, {Ptak}, \& {Roberts}}]{Lehmer2010}
{Lehmer}, B.~D., {Alexander}, D.~M., {Bauer}, F.~E., {et~al.} 2010, \apj, 724,
  559

\bibitem[{{Lehmer} {et~al.}(2016){Lehmer}, {Basu-Zych}, {Mineo}, {Brandt},
  {Eufrasio}, {Fragos}, {Hornschemeier}, {Luo}, {Xue}, {Bauer}, {Gilfanov},
  {Ranalli}, {Schneider}, {Shemmer}, {Tozzi}, {Trump}, {Vignali}, {Wang},
  {Yukita}, \& {Zezas}}]{Lehmer2016}
{Lehmer}, B.~D., {Basu-Zych}, A.~R., {Mineo}, S., {et~al.} 2016, \apj, 825, 7

\bibitem[{{Lehmer} {et~al.}(2012){Lehmer}, {Xue}, {Brandt}, {Alexander},
  {Bauer}, {Brusa}, {Comastri}, {Gilli}, {Hornschemeier}, {Luo}, {Paolillo},
  {Ptak}, {Shemmer}, {Schneider}, {Tozzi}, \& {Vignali}}]{Lehmer2012}
{Lehmer}, B.~D., {Xue}, Y.~Q., {Brandt}, W.~N., {et~al.} 2012, \apj, 752, 46

\bibitem[{{Leitherer} {et~al.}(2002){Leitherer}, {Calzetti}, \&
  {Martins}}]{Leitherer2002}
{Leitherer}, C., {Calzetti}, D., \& {Martins}, L.~P. 2002, \apj, 574, 114

\bibitem[{{Linden} {et~al.}(2010){Linden}, {Kalogera}, {Sepinsky}, {Prestwich},
  {Zezas}, \& {Gallagher}}]{Linden2010}
{Linden}, T., {Kalogera}, V., {Sepinsky}, J.~F., {et~al.} 2010, \apj, 725, 1984

\bibitem[{{Liu} {et~al.}(2022){Liu}, {Buchner}, {Nandra}, {Merloni}, {Dwelly},
  {Sanders}, {Salvato}, {Arcodia}, {Brusa}, {Wolf}, {Georgakakis}, {Boller},
  {Krumpe}, {Lamer}, {Waddell}, {Urrutia}, {Schwope}, {Robrade}, {Wilms},
  {Dauser}, {Comparat}, {Toba}, {Ichikawa}, {Iwasawa}, {Shen}, \&
  {Medel}}]{Liu2022}
{Liu}, T., {Buchner}, J., {Nandra}, K., {et~al.} 2022, \aap, 661, A5

\bibitem[{{Lo Faro} {et~al.}(2017){Lo Faro}, {Buat}, {Roehlly},
  {Alvarez-Marquez}, {Burgarella}, {Silva}, \& {Efstathiou}}]{LoFaro2017}
{Lo Faro}, B., {Buat}, V., {Roehlly}, Y., {et~al.} 2017, \mnras, 472, 1372

\bibitem[{{Luo} {et~al.}(2017){Luo}, {Brandt}, {Xue}, {Lehmer}, {Alexander},
  {Bauer}, {Vito}, {Yang}, {Basu-Zych}, {Comastri}, {Gilli}, {Gu},
  {Hornschemeier}, {Koekemoer}, {Liu}, {Mainieri}, {Paolillo}, {Ranalli},
  {Rosati}, {Schneider}, {Shemmer}, {Smail}, {Sun}, {Tozzi}, {Vignali}, \&
  {Wang}}]{Luo2017}
{Luo}, B., {Brandt}, W.~N., {Xue}, Y.~Q., {et~al.} 2017, \apjs, 228, 2

\bibitem[{{Ma{\l}ek} {et~al.}(2018){Ma{\l}ek}, {Buat}, {Roehlly}, {Burgarella},
  {Hurley}, {Shirley}, {Duncan}, {Efstathiou}, {Papadopoulos}, {Vaccari},
  {Farrah}, {Marchetti}, \& {Oliver}}]{Malek2018}
{Ma{\l}ek}, K., {Buat}, V., {Roehlly}, Y., {et~al.} 2018, \aap, 620, A50

\bibitem[{{McMahon} {et~al.}(2013){McMahon}, {Banerji}, {Gonzalez}, {Koposov},
  {Bejar}, {Lodieu}, {Rebolo}, \& {VHS Collaboration}}]{McMahon2013}
{McMahon}, R.~G., {Banerji}, M., {Gonzalez}, E., {et~al.} 2013, The Messenger,
  154, 35

\bibitem[{{Meisner} {et~al.}(2019){Meisner}, {Lang}, {Schlafly}, \&
  {Schlegel}}]{Meisner2019}
{Meisner}, A.~M., {Lang}, D., {Schlafly}, E.~F., \& {Schlegel}, D.~J. 2019,
  \pasp, 131, 124504

\bibitem[{{Merloni} {et~al.}(2020){Merloni}, {Nandra}, \&
  {Predehl}}]{Merloni2020}
{Merloni}, A., {Nandra}, K., \& {Predehl}, P. 2020, Nature Astronomy, 4, 634

\bibitem[{{Merloni} {et~al.}(2012){Merloni}, {Predehl}, {Becker},
  {B{\"o}hringer}, {Boller}, {Brunner}, {Brusa}, {Dennerl}, {Freyberg},
  {Friedrich}, {Georgakakis}, {Haberl}, {Hasinger}, {Meidinger}, {Mohr},
  {Nandra}, {Rau}, {Reiprich}, {Robrade}, {Salvato}, {Santangelo}, {Sasaki},
  {Schwope}, {Wilms}, \& {German eROSITA Consortium}}]{Merloni2012}
{Merloni}, A., {Predehl}, P., {Becker}, W., {et~al.} 2012, arXiv e-prints,
  arXiv:1209.3114

\bibitem[{{Mewe} {et~al.}(1986){Mewe}, {Lemen}, \& {van den Oord}}]{Mewe1986}
{Mewe}, R., {Lemen}, J.~R., \& {van den Oord}, G.~H.~J. 1986, \aaps, 65, 511

\bibitem[{{Mineo} {et~al.}(2014){Mineo}, {Gilfanov}, {Lehmer}, {Morrison}, \&
  {Sunyaev}}]{Mineo2014}
{Mineo}, S., {Gilfanov}, M., {Lehmer}, B.~D., {Morrison}, G.~E., \& {Sunyaev},
  R. 2014, \mnras, 437, 1698

\bibitem[{{Mineo} {et~al.}(2012){Mineo}, {Gilfanov}, \& {Sunyaev}}]{Mineo2012}
{Mineo}, S., {Gilfanov}, M., \& {Sunyaev}, R. 2012, \mnras, 419, 2095

\bibitem[{{Mountrichas} {et~al.}(2022){Mountrichas}, {Buat}, {Yang}, {Boquien},
  {Burgarella}, {Ciesla}, {Malek}, \& {Shirley}}]{Mountrichas2022}
{Mountrichas}, G., {Buat}, V., {Yang}, G., {et~al.} 2022, \aap, 663, A130

\bibitem[{{Netzer}(1987)}]{Netzer1987}
{Netzer}, H. 1987, \mnras, 225, 55

\bibitem[{{Noll} {et~al.}(2009){Noll}, {Burgarella}, {Giovannoli}, {Buat},
  {Marcillac}, \& {Mu{\~n}oz-Mateos}}]{Noll2009}
{Noll}, S., {Burgarella}, D., {Giovannoli}, E., {et~al.} 2009, \aap, 507, 1793

\bibitem[{{Pellegrini}(1994)}]{Pellegrini1994}
{Pellegrini}, S. 1994, \aap, 292, 395

\bibitem[{{Poglitsch} {et~al.}(2010){Poglitsch}, {Waelkens}, {Geis},
  {Feuchtgruber}, {Vandenbussche}, {Rodriguez}, {Krause}, {Renotte}, {van
  Hoof}, {Saraceno}, {Cepa}, {Kerschbaum}, {Agn{\`e}se}, {Ali}, {Altieri},
  {Andreani}, {Augueres}, {Balog}, {Barl}, {Bauer}, {Belbachir}, {Benedettini},
  {Billot}, {Boulade}, {Bischof}, {Blommaert}, {Callut}, {Cara}, {Cerulli},
  {Cesarsky}, {Contursi}, {Creten}, {De Meester}, {Doublier}, {Doumayrou},
  {Duband }, {Exter}, {Genzel}, {Gillis}, {Gr{\"o}zinger}, {Henning},
  {Herreros}, {Huygen}, {Inguscio}, {Jakob}, {Jamar}, {Jean}, {de Jong},
  {Katterloher}, {Kiss}, {Klaas}, {Lemke}, {Lutz}, {Madden}, {Marquet},
  {Martignac}, {Mazy}, {Merken}, {Montfort}, {Morbidelli}, {M{\"u}ller},
  {Nielbock}, {Okumura}, {Orfei}, {Ottensamer}, {Pezzuto}, {Popesso},
  {Putzeys}, {Regibo}, {Reveret}, {Royer}, {Sauvage}, {Schreiber}, {Stegmaier},
  {Schmitt}, {Schubert}, {Sturm}, {Thiel}, {Tofani}, {Vavrek}, {Wetzstein},
  {Wieprecht}, \& {Wiezorrek}}]{Poglitsch2010}
{Poglitsch}, A., {Waelkens}, C., {Geis}, N., {et~al.} 2010, \aap, 518, L2

\bibitem[{{Predehl} {et~al.}(2021){Predehl}, {Andritschke}, {Arefiev},
  {Babyshkin}, {Batanov}, {Becker}, {B{\"o}hringer}, {Bogomolov}, {Boller},
  {Borm}, {Bornemann}, {Br{\"a}uninger}, {Br{\"u}ggen}, {Brunner}, {Brusa},
  {Bulbul}, {Buntov}, {Burwitz}, {Burkert}, {Clerc}, {Churazov}, {Coutinho},
  {Dauser}, {Dennerl}, {Doroshenko}, {Eder}, {Emberger}, {Eraerds},
  {Finoguenov}, {Freyberg}, {Friedrich}, {Friedrich}, {F{\"u}rmetz},
  {Georgakakis}, {Gilfanov}, {Granato}, {Grossberger}, {Gueguen}, {Gureev},
  {Haberl}, {H{\"a}lker}, {Hartner}, {Hasinger}, {Huber}, {Ji}, {Kienlin},
  {Kink}, {Korotkov}, {Kreykenbohm}, {Lamer}, {Lomakin}, {Lapshov}, {Liu},
  {Maitra}, {Meidinger}, {Menz}, {Merloni}, {Mernik}, {Mican}, {Mohr},
  {M{\"u}ller}, {Nandra}, {Nazarov}, {Pacaud}, {Pavlinsky}, {Perinati},
  {Pfeffermann}, {Pietschner}, {Ramos-Ceja}, {Rau}, {Reiffers}, {Reiprich},
  {Robrade}, {Salvato}, {Sanders}, {Santangelo}, {Sasaki}, {Scheuerle},
  {Schmid}, {Schmitt}, {Schwope}, {Shirshakov}, {Steinmetz}, {Stewart},
  {Str{\"u}der}, {Sunyaev}, {Tenzer}, {Tiedemann}, {Tr{\"u}mper}, {Voron},
  {Weber}, {Wilms}, \& {Yaroshenko}}]{Predehl2021}
{Predehl}, P., {Andritschke}, R., {Arefiev}, V., {et~al.} 2021, \aap, 647, A1

\bibitem[{{Ramos Padilla} {et~al.}(2022){Ramos Padilla}, {Wang}, {Ma{\l}ek},
  {Efstathiou}, \& {Yang}}]{Padilla2022}
{Ramos Padilla}, A.~F., {Wang}, L., {Ma{\l}ek}, K., {Efstathiou}, A., \&
  {Yang}, G. 2022, \mnras, 510, 687

\bibitem[{{Riccio} {et~al.}(2021){Riccio}, {Ma{\l}ek}, {Nanni}, {Boquien},
  {Buat}, {Burgarella}, {Donevski}, {Hamed}, {Hurley}, {Shirley}, \&
  {Pollo}}]{Riccio2021}
{Riccio}, G., {Ma{\l}ek}, K., {Nanni}, A., {et~al.} 2021, \aap, 653, A107

\bibitem[{{Ruiz} {et~al.}(2018){Ruiz}, {Corral}, {Mountrichas}, \&
  {Georgantopoulos}}]{Ruiz2018}
{Ruiz}, A., {Corral}, A., {Mountrichas}, G., \& {Georgantopoulos}, I. 2018,
  \aap, 618, A52

\bibitem[{{Salvato} {et~al.}(2018){Salvato}, {Buchner}, {Budav{\'a}ri},
  {Dwelly}, {Merloni}, {Brusa}, {Rau}, {Fotopoulou}, \& {Nandra}}]{Salvato2018}
{Salvato}, M., {Buchner}, J., {Budav{\'a}ri}, T., {et~al.} 2018, \mnras, 473,
  4937

\bibitem[{{Salvato} {et~al.}(2022){Salvato}, {Wolf}, {Dwelly}, {Georgakakis},
  {Brusa}, {Merloni}, {Liu}, {Toba}, {Nandra}, {Lamer}, {Buchner}, {Schneider},
  {Freund}, {Rau}, {Schwope}, {Nishizawa}, {Klein}, {Arcodia}, {Comparat},
  {Musiimenta}, {Nagao}, {Brunner}, {Malyali}, {Finoguenov}, {Anderson},
  {Shen}, {Ibarra-Medel}, {Trump}, {Brandt}, {Urry}, {Rivera}, {Krumpe},
  {Urrutia}, {Miyaji}, {Ichikawa}, {Schneider}, {Fresco}, {Boller}, {Haase},
  {Brownstein}, {Lane}, {Bizyaev}, \& {Nitschelm}}]{Salvato2022}
{Salvato}, M., {Wolf}, J., {Dwelly}, T., {et~al.} 2022, \aap, 661, A3

\bibitem[{{Shirley} {et~al.}(2021){Shirley}, {Duncan}, {Campos Varillas},
  {Hurley}, {Ma{\l}ek}, {Roehlly}, {Smith}, {Aussel}, {Bakx}, {Buat},
  {Burgarella}, {Christopher}, {Duivenvoorden}, {Eales}, {Efstathiou},
  {Gonz{\'a}lez Solares}, {Griffin}, {Jarvis}, {Faro}, {Marchetti}, {McCheyne},
  {Papadopoulos}, {Penner}, {Pons}, {Prescott}, {Rigby}, {Rottgering},
  {Saxena}, {Scudder}, {Vaccari}, {Wang}, \& {Oliver}}]{Shirley2021}
{Shirley}, R., {Duncan}, K., {Campos Varillas}, M.~C., {et~al.} 2021, \mnras,
  507, 129

\bibitem[{{Shtykovskiy} \& {Gilfanov}(2007)}]{Gilfanov2007}
{Shtykovskiy}, P.~E. \& {Gilfanov}, M.~R. 2007, Astronomy Letters, 33, 437

\bibitem[{{Stalevski} {et~al.}(2012){Stalevski}, {Fritz}, {Baes}, {Nakos}, \&
  {Popovi{\'c}}}]{Stalevski2012}
{Stalevski}, M., {Fritz}, J., {Baes}, M., {Nakos}, T., \& {Popovi{\'c}},
  L.~{\v{C}}. 2012, \mnras, 420, 2756

\bibitem[{{Stalevski} {et~al.}(2016){Stalevski}, {Ricci}, {Ueda}, {Lira},
  {Fritz}, \& {Baes}}]{Stalevski2016}
{Stalevski}, M., {Ricci}, C., {Ueda}, Y., {et~al.} 2016, \mnras, 458, 2288

\bibitem[{{Suleiman} {et~al.}(2022){Suleiman}, {Noboriguchi}, {Toba},
  {Bal{\'a}zs}, {Burgarella}, {Kov{\'a}cs}, {Marton}, {Talafha}, {Frey}, \&
  {T{\'o}th}}]{Suleiman2022}
{Suleiman}, N., {Noboriguchi}, A., {Toba}, Y., {et~al.} 2022, \pasj, 74, 1157

\bibitem[{{Sunyaev} {et~al.}(2021){Sunyaev}, {Arefiev}, {Babyshkin},
  {Bogomolov}, {Borisov}, {Buntov}, {Brunner}, {Burenin}, {Churazov},
  {Coutinho}, {Eder}, {Eismont}, {Freyberg}, {Gilfanov}, {Gureyev}, {Hasinger},
  {Khabibullin}, {Kolmykov}, {Komovkin}, {Krivonos}, {Lapshov}, {Levin},
  {Lomakin}, {Lutovinov}, {Medvedev}, {Merloni}, {Mernik}, {Mikhailov},
  {Molodtsov}, {Mzhelsky}, {M{\"u}ller}, {Nandra}, {Nazarov}, {Pavlinsky},
  {Poghodin}, {Predehl}, {Robrade}, {Sazonov}, {Scheuerle}, {Shirshakov},
  {Tkachenko}, \& {Voron}}]{Sunyaev2021}
{Sunyaev}, R., {Arefiev}, V., {Babyshkin}, V., {et~al.} 2021, \aap, 656, A132

\bibitem[{{Sutherland} \& {Saunders}(1992)}]{Sutherland1992}
{Sutherland}, W. \& {Saunders}, W. 1992, \mnras, 259, 413

\bibitem[{{Terlevich} \& {Melnick}(1985)}]{Terlevitch}
{Terlevich}, R. \& {Melnick}, J. 1985, \mnras, 213, 841

\bibitem[{{Torbaniuk} {et~al.}(2021){Torbaniuk}, {Paolillo}, {Carrera},
  {Cavuoti}, {Vignali}, {Longo}, \& {Aird}}]{Torbaniuk2021}
{Torbaniuk}, O., {Paolillo}, M., {Carrera}, F., {et~al.} 2021, \mnras, 506,
  2619

\bibitem[{{Tremonti} {et~al.}(2004{\natexlab{a}}){Tremonti}, {Heckman},
  {Kauffmann}, {Brinchmann}, {Charlot}, {White}, {Seibert}, {Peng}, {Schlegel},
  {Uomoto}, {Fukugita}, \& {Brinkmann}}]{Tremonti}
{Tremonti}, C.~A., {Heckman}, T.~M., {Kauffmann}, G., {et~al.}
  2004{\natexlab{a}}, \apj, 613, 898

\bibitem[{{Tremonti} {et~al.}(2004{\natexlab{b}}){Tremonti}, {Heckman},
  {Kauffmann}, {Brinchmann}, {Charlot}, {White}, {Seibert}, {Peng}, {Schlegel},
  {Uomoto}, {Fukugita}, \& {Brinkmann}}]{Tremonti2004}
{Tremonti}, C.~A., {Heckman}, T.~M., {Kauffmann}, G., {et~al.}
  2004{\natexlab{b}}, \apj, 613, 898

\bibitem[{{T{\"u}llmann} {et~al.}(2006){T{\"u}llmann}, {Pietsch}, {Rossa},
  {Breitschwerdt}, \& {Dettmar}}]{Tullmann2006}
{T{\"u}llmann}, R., {Pietsch}, W., {Rossa}, J., {Breitschwerdt}, D., \&
  {Dettmar}, R.~J. 2006, \aap, 448, 43

\bibitem[{{Verbunt} \& {van den Heuvel}(1995)}]{Verbunt1995}
{Verbunt}, F. \& {van den Heuvel}, E.~P.~J. 1995, in X-ray Binaries, 457--494

\bibitem[{{Vulic} {et~al.}(2022){Vulic}, {Hornschemeier}, {Haberl},
  {Basu-Zych}, {Kyritsis}, {Zezas}, {Salvato}, {Ptak}, {Bogdan}, {Kovlakas},
  {Wilms}, {Sasaki}, {Liu}, {Merloni}, {Dwelly}, {Brunner}, {Lamer}, {Maitra},
  {Nandra}, \& {Santangelo}}]{Vulic2022}
{Vulic}, N., {Hornschemeier}, A.~E., {Haberl}, F., {et~al.} 2022, \aap, 661,
  A16

\bibitem[{{Xue} {et~al.}(2011){Xue}, {Luo}, {Brandt}, {Bauer}, {Lehmer},
  {Broos}, {Schneider}, {Alexander}, {Brusa}, {Comastri}, {Fabian}, {Gilli},
  {Hasinger}, {Hornschemeier}, {Koekemoer}, {Liu}, {Mainieri}, {Paolillo},
  {Rafferty}, {Rosati}, {Shemmer}, {Silverman}, {Smail}, {Tozzi}, \&
  {Vignali}}]{Xue2011}
{Xue}, Y.~Q., {Luo}, B., {Brandt}, W.~N., {et~al.} 2011, \apjs, 195, 10

\bibitem[{{Yang} {et~al.}(2022){Yang}, {Boquien}, {Brandt}, {Buat},
  {Burgarella}, {Ciesla}, {Lehmer}, {Ma{\l}ek}, {Mountrichas}, {Papovich},
  {Pons}, {Stalevski}, {Theul{\'e}}, \& {Zhu}}]{Yang2021}
{Yang}, G., {Boquien}, M., {Brandt}, W.~N., {et~al.} 2022, \apj, 927, 192

\bibitem[{{Yang} {et~al.}(2020){Yang}, {Boquien}, {Buat}, {Burgarella},
  {Ciesla}, {Duras}, {Stalevski}, {Brandt}, \& {Papovich}}]{Guang2020}
{Yang}, G., {Boquien}, M., {Buat}, V., {et~al.} 2020, \mnras, 491, 740

\bibitem[{{Yuan} {et~al.}(2018){Yuan}, {Argudo-Fern{\'a}ndez}, {Shen}, {Hao},
  {Jiang}, {Yin}, {Boquien}, \& {Lin}}]{Yuan2018}
{Yuan}, F.-T., {Argudo-Fern{\'a}ndez}, M., {Shen}, S., {et~al.} 2018, \aap,
  613, A13

\end{thebibliography}
\end{document}